\documentclass[submission]{eptcs}
\usepackage{breakurl}             
\usepackage{underscore}           
\usepackage{macros}
\usepackage{subfigure}
\usepackage{amsfonts}
\usepackage{graphicx}
\usepackage{epstopdf}
\usepackage{float}
\usepackage{mathtools}
\usepackage[table,xcdraw]{xcolor}
\usepackage{listings}
\usepackage{latexsym}
\usepackage{amssymb}
\newcommand{\remove}[1]{} 
\usepackage{parallel,enumitem}
\usepackage[subfigure]{tocloft}
\usepackage{amsfonts}
\usepackage{graphicx}

\newtheorem{definition}{Definition}
\newtheorem{lemma}{Lemma}
\newtheorem{theorem}{Theorem}
\newtheorem{example}{Example}
\newtheorem{proposition}{Proposition}

\newtheorem{corollary}{Corollary}

\title{Reversible Computation in Petri Nets}
\author{Anna Philippou
\institute{Department of Computer Science \\
University of Cyprus}
\email{annap@cs.ucy.ac.cy}
\and
 Kyriaki Psara
\institute{Department of Computer Science \\
University of Cyprus}
\email{ kpsara01@cs.ucy.ac.cy }
}
\def\titlerunning{Reversible Computation in Petri nets}
\def\authorrunning{A. Philippou \& K. Psara}
\begin{document}
\maketitle

\begin{abstract}
Reversible computation is an unconventional form of computing where any executed sequence of operations
can be executed in reverse at any point during computation. It has recently been attracting increasing attention
in various research communities as on the one hand it promises low-power computation and on the other hand
it is inherent or of interest in a variety of applications. In this paper, we propose a reversible approach to Petri nets by introducing 
machinery and associated operational semantics to tackle the challenges of the three main forms of reversibility, namely, 
backtracking, causal reversing and out-of-causal-order reversing. Our proposal concerns a variation of Petri nets where 
tokens are persistent and are distinguished from each other by an identity which allows for transitions to be reversed 
spontaneously in or out of causal order. Our design decisions are influenced by applications in biochemistry but the 
methodology can be applied to a wide range of problems that feature reversibility. In particular, to demonstrate the 
applicability of our approach
we use an example of a biochemical system and an example of a  transaction-processing system both of which naturally 
embed reversible behaviour.
\end{abstract}

\pagestyle{pla5 in}
\pagenumbering{arabic}

\section{Introduction\vspace{-.5em}}\label{sec:Introduction}

Reversible computation is an unconventional form of computing where computation can
be executed in forward direction as effortlessly as it can be executed in backward direction. 
Any sequence of operations carried out by a system can be subsequently executed reversibly
allowing the system to  retrieve any previous state at any point during computation. 
The motivation for reversible computing began with  Landauer's observation that only irreversible computation
generates heat \cite{Landauer} spawning a strong line of research  towards the creation of reversible logic
gates and circuits. 
Subsequently, motivation for studying reversibility has stemmed from a wide variety of applications
which  naturally embed reversible behaviour.  These include biological processes where computation may be carried
out in forward or backward direction~\cite{ERK,LocalRev}, and
the field of system reliability
where reversibility can be used as means of recovering from failures~\cite{TransactionsRCCS,LaneseLMSS13}.

Several subcategories of reversible computation have been identified and studied in the past years.
These include \emph{backtracking} and the more general form of reversibility referred to as \emph{causal-order reversibility} according to which
a transition can be undone only if all its effects, if any, have been undone beforehand. Attention has also
turned towards reversing in \emph{out-of-causal order}, a form of reversing featured most notably in biochemical systems. These concepts have been studied within a variety of formalisms. To begin with, a large
amount of work has focused on providing a formal understanding of reversibility within process calculi. The first reversible process calculus dates back to 2004 when Danos and Krivine proposed 
RCCS~\cite{RCCS}, a causal-consistent reversible extension of CCS, further developed in \cite{TransactionsRCCS,BiologyCCSR}. 
Soon after, Phillips and Ulidowski proposed a general method for reversing process calculi, with their proposal of CCSK
 being a special instance of the methodology~\cite{Algebraic}. 
 Constructs for controlling reversibility were also proposed 
in reversible extensions of the $\pi$-calculus in~\cite{LaneseLMSS13,LaneseMS16}.
Most recently, the study of out-of-causal-order reversibility continued with the introduction of a new operator for modelling
local reversibility  
in \cite{LocalRev}. Furthermore, reversible computation was studied within event structures 
in~\cite{ConRev}. The modelling of bonding within reversible processes and event structures was also considered
in~\cite{Bonding} whereas a reversible computational calculus for modelling chemical systems composed of signals 
and gates was  proposed in~\cite{CardelliL11}.

Petri nets (PNs) are a graphical mathematical language that can be used for the specification and
analysis of discrete event systems. They are associated with 
a rich mathematical theory and a variety of tools, and they have been used extensively
for modelling and reasoning about a wide range of applications.
A property studied in the context of Petri nets is that of \emph{Petri net reversibility}
which describes the ability of  a system to return to the initial state from any reachable state. 
This, however, is in contrast to the notion of reversible computation as discussed above
where the intention is not to return to a state via arbitrary execution but to reverse
the effect of already executed transitions.

More recently the first study of reversible computation within Petri nets was proposed in~\cite{PetriNets,BoundedPNs}. In these works, the
authors investigate the effects of adding \emph{reversed} versions of selected transitions in a Petri net, where these
transitions are obtained by reversing the directions of a transition's arcs. They then explore decidability
problems regarding reachability and coverability in the resulting Petri nets. Related to our study of causal reversibility
is also work carried out regarding causality in Petri nets. We mention~\cite{GlabbeekGS11,Glabbeek05,Glabbeek}
where the authors explore causal semantics in Petri nets by utilising the individual token interpretation to
distinguish tokens as well as the collective token interpretation. 
\vspace{0.05in}

\noindent{\bf Contribution.}
In this work we set out to study reversible computation in the context of Petri nets and in particular to explore the modelling of the  main strategies for reversing computation. Our aim is
to address the challenges of capturing the notions of \emph{backtracking}, 
\emph{causal reversibility} and \emph{out-of-causal-order reversibility} within the Petri net framework, thus proposing a novel, graphical methodology for studying reversibility in a model where transitions can be taken in either direction.  

Our proposal is motivated by applications from biochemistry where out-of-causal-order reversibility is inherent, and it supports all the forms of reversibility that have been discussed above.
In particular, we consider a subclass of PNs which are acyclic and where tokens are persistent.
We prove that the amount of flexibility allowed in causal reversibility indeed yields a causally consistent semantics. We also demonstrate 
that out-of-causal  reversibility
is  able to create new states unreachable by forward-only execution which, nonetheless, respect causality with regard to connected components
of tokens. 

\vspace{0.05in}
\noindent{\bf Paper organisation.}
In the next section we give an overview of the different types of reversibility and their characteristics and
we discuss the challenges of modelling reversibility in the context of Petri nets. In Section 3 we
introduce the formalism of  Reversible Petri nets, and, in Section 4, we present semantics for
the models that capture backtracking, causal and out-of-causal-order reversibility. We illustrate the framework with 
an example inspired by long-running transactions.
Section 5 concludes the paper. Missing proofs can be found in an appendix.

\section{Forms of Reversibility and Petri Nets}\label{sec:Forms of Reversibility}

Even though reversing computational processes in concurrent and distributed systems has 
many promising applications, it also has many technical and conceptual challenges. One of 
the most important questions that arise regards the strategy to be applied when going backwards. 
Several approaches 
have been explored in the literature over the last decade which differ in the order
in which steps are taken backwards. The most prominent of these are \emph{backtracking}, 
\emph{causal reversibility}, and \emph{out-of-causal-order reversibility}. 

{\em Backtracking} is the process of rewinding one's computation trace, that is, computational steps 
are undone in the exact inverse order to the one in which they have occurred. 
This form of reversing ensures that at any state in a computation there is at most one predecessor 
state. 
In the context of concurrent systems, this form of reversibility can be thought of as overly restrictive since, undoing moves only
in the order in which they were taken, induces fake causal dependencies on backward sequences of actions: 
actions, which could have been reversed in any order are forced to be undone in the precise order in which they occurred.
		
A second approach to reversibility, named {\em causal reversibility}, relaxes the rigidity of backtracking. It allows a more flexible form of reversibility by allowing events to reverse in an arbitrary order assuming that
they respect the causal dependencies that hold between them. Thus, in the context of
causal reversibility, reversing does not have to follow the exact inverse order for independent events as long 
as caused actions, also known as effects, are undone before the actions that have caused 
them. 

\begin{figure}
\centering
\subfigure{\includegraphics[width=8cm]{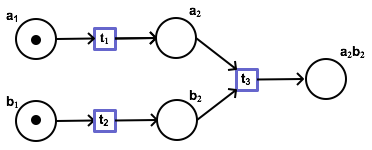}}
\caption{ Causal Reversibility}
\label{causal}
\end{figure}

For example consider the Petri net in Figure \ref{causal}.  We may observe
that transitions $t_1$ and $t_2$ are independent from each other, as they may be taken in any
order, and they are both prerequisites for transition $t_3$. 
Backtracking the sequence of transitions  $\langle t_1, t_2,t_3\rangle$ would require that
 the three transitions should be reversed in exactly the reverse order, i.e. 
 $\langle t_3, t_2,t_1\rangle$. Instead, causal flexibility 
allows the inverse computation to rewind $t_3$ and then $t_1$ and $t_2$ in any order (but never
$t_1$ or $t_2$  before $t_3$).


Both backtracking and causal reversing are cause-respecting. There are, however, many 
real-life examples where undoing things in an {\em out-of-causal} order is either inherent or
could be beneficial. In fact, this form of undoing plays a vital role on mechanisms driving long-running transactions and biochemical reactions. 
Consider every state of the execution to be a result of a series of actions that have causally contributed 
to the existence of the current state. If the actions were to be reversed in a causally-respecting manner 
then we would only be able to move back and forth through previously visited states. Therefore, one
might wish to apply out-of-order reversibility in order to create fresh alternatives of current states that 
were formerly inaccessible by any forward-only execution path.

Since out-of-order reversibility  contradicts program order by violating the laws of causality, it comes with its 
own peculiarities that need to be taken into consideration while designing reversible systems. To appreciate
these peculiarities and obtain insights towards our approach on addressing reversibility within Petri nets, we
will  use a standard example from the literature, namely the process of catalysis from biochemistry. 

\begin{figure}
\centering
\subfigure{\includegraphics[width=8cm]{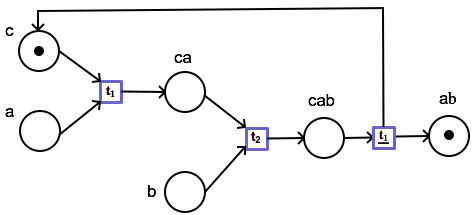}}
\caption{ Catalysis in classic Petri nets}
\label{catalyst2}
\end{figure}

Consider the catalyst $c$ that helps the otherwise inactive molecules $a$  and $b$ to bond. The
process followed to achieve catalysis is element $c$ bonding with $a$ which then enables the bonding 
between $a$ and $b$. Next, the catalyst is no longer needed and its bond to
the other two molecules is released. A Petri net model of this process is illustrated in Figure~\ref{catalyst2}.
The Petri net executes transition $t_1$ via which the bond $ca$ is created, followed by
action $t_2$ to produce $cab$. Finally, action
$\underline{t_1}$ ``reverses'' the bond between $a$ and $c$, yielding $ab$ and releasing $c$. (The figure portrays the final state of the execution assuming that initially exactly one token existed in places $a$, $b$, and $c$.)

This example illustrates that Petri nets are not reversible by nature, in the sense that every transition cannot be executed
in both directions. Therefore an inverse action, (e.g. transition $\underline{t_1}$  for
undoing the effect of transition $t_1$, namely bond $ca$) needs to be added as a supplementary forward transition for achieving the undoing of a previous action. 
This explicit approach of modelling reversibility can prove cumbersome in systems that express multiple reversible
patterns of execution, resulting in larger and more complex systems. Furthermore, it
 fails to capture reversibility as a mode of computation.
The intention of our work is to study an approach for modelling reversible computation that does not require the addition of new, reversed transitions but
instead allows to execute transitions in both the forward as well as the backward direction, and, thereby, explore the theory
of reversible computation within Petri nets.

However, when attempting to model the catalysis example while executing transitions in both
the forward and the backward directions, we may observe a number of obstacles. At an abstract level, the behaviour of the system should exhibit a sequence of three transitions: execution of $t_1$ and $t_2$, followed by the reversal of transition $t_1$. 
The reversal of transition $t_1$ should implement the release of $c$ from the bond $cab$ and make
it available for further instantiations of transition $t_1$, if needed, while the bond $ab$ should remain in place.
This implies that a reversing Petri net model should provide resources
$a$, $b$ and $c$ as well as $ca$, $cab$ and $ab$ and implement the reversal of action $t_1$ as the
transformation of resource $cab$ into $c$ and $ab$. Note that resource $ab$ is inaccessible during the forward execution of transitions $t_1$ and $t_2$ and only materialises after the reversal of transition $t_1$,
i.e. only once the bond between $a$ and $c$ is broken. Given the static nature of a Petri net, this suggests that
resources such as $ab$ should be  represented at the token
level (as opposed to the place level). As a result, the concept of token individuality is of particular relevance to
reversible computation in Petri nets while other constructs/functions at token level are needed to capture the effect and reversal
of a transition.

Indeed, reversing a transition in an out-of-causal order may imply that while some of the effects of the transition
can be reversed (e.g. the release of the catalyst back to the initial state), others must be retained due to computation that succeeded the forward execution of the next transition (e.g. token $a$ cannot be released during the reversal of $t_1$ since
it has bonded with $b$ in transition $t_2$). This latter point is especially challenging since it requires to determine a model in a precise manner so as to identify which effects are allowed to be ``undone'' when reversing a transition. We will now proceed to discuss our design decisions towards our proposal for a reversing Petri net framework. 

\subsection{Design Choices}

\begin{figure}
\centering
\subfigure{\includegraphics[width=6.7cm]{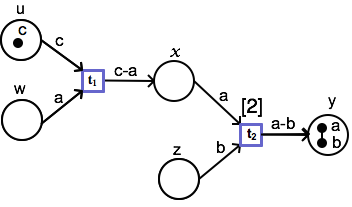}}
\caption{ Catalysis in reversing Petri nets}
\label{catalyst}
\end{figure}

As highlighted by the catalyst example, reversing transitions in a Petri net model requires close monitoring of
token manipulation within a net and clear enunciation of the effects of a transition. 
In particular, undoing a transition may result in some tokens to return to the in-places of the transition, while others will
not do so unless and/or until further transitions in which the tokens were involved are additionally reversed.
In order to achieve this one we would need to distinguish each individual token
along with sufficient information regarding its causal path, i.e., the places and transitions
it has traversed before reaching its current state. This requirement is rather demanding and would
require special care in the case where transitions consume 
multiple tokens but release only a subset of the tokens or even a set of new tokens.

In our approach we employ the notion of token individuality but, instead of maintaining extensive histories for recording
the precise evolution of each token through transitions and places, we have opted to employ a novel approach inspired
by out-of-causal reversibility in biochemistry as well as approaches from related literature~\cite{Bonding}. The resulting 
framework is light in
the sense that no memory needs to be stored per token to retrieve its causal path
while enabling reversible semantics for the three main types of reversibility. Furthermore, while inspired by biochemistry, the framework
can be applied in other contexts of reversible systems: for instance in Section 4.4 we provide an example of 
reversible transactions.

Specifically, our approach similarly to Coloured Petri Nets allows the observation of token evolution through transition firing. We introduce two notions that intuitively capture tokens and their history: the notion of \emph{base} and a new type of tokens called \emph{bonds}. A base is a persistent type of token which cannot be
consumed and therefore preserves its individuality through various transitions. For a transition to fire, the incoming arcs identify the required tokens/bonds and the outgoing arcs may create new bonds or transfer already existing tokens/bonds along the places of a PN.  
Therefore, the effect of a transition is the creation of new \emph{bonds} between the tokens it takes as input and the reversal of such a transition involves undoing the respective bonds. 
 In other words, a bonded token is a  coalition of bases connected via bonds into a structure located in a place, thus presenting a single entity while individually representing each base/bond. 
  
Based on these ideas, we may describe the catalyst example in our proposed framework as shown in Figure~\ref{catalyst}. 
In this new setting $a$ and $c$ are bases which during transition $t_1$ are connected via a bond into place $x$, while 
transition $t_2$ brings into place a new bond between $a$ and $b$.  In Figure~\ref{catalyst} we may see the state that arises after execution of $t_1$ and $t_2$ and the reversal of transition $t_1$. In this state, base $c$ has returned to its initial place $u$ whereas bond $a-b$ has remained in place $y$. A thorough explanation of the notation is given in the next section.




The next design decision in our methodology is that we offer the flexibility for transitions to take as input 
a subset of the bonds and bases, thus
denoting exactly the entities that are necessary for the firing of a transition by labelling the incoming arcs with necessary tokens. 
If further bases/bonds exist in a place but are not featured in the incoming arcs to a transition, they do not
necessarily need to exist for the transition to fire. Such absence of tokens
may occur due to a previous reversal of a transition
that resulted in tokens backtracking along the net. However, if such bases/bonds do exist in an in-place, 
the transition should output the respective tokens in its output places, thus preserving the bonds. 

The final design decision of our framework stems from the need to identify at each point in time the
history of execution, a necessary aspect for all forms of reversibility: observing a state of a Petri net,
 with tokens scattered along its places, due to the nondeterministic nature of Petri nets,
  it is not possible to discern the history that led to the specific 
 state and consequently the precise transitions that can be undone. While some aspects of this history 
 are made available by the persistence of bases and the creation of bonds, due to nondeterminism we 
 need to distinguish the cause of a certain state among the several possible alternatives. To achieve 
 this, we associate transitions with a history where we store keys in increasing order each time an 
instance of the transition is executed. This allows to backtrack computation 
as well as to extract the causes of bonds as
needed in causal and out-of-causal-order reversibility. These identifiers can be removed from the history 
once the respective transition has been reversed. 

As a final note, we point out that in what follows we make the assumption that we may have no more than one token for each base in a Reversible Petri net model and that models are acyclic. The assumptions enable a clearer enunciation and presentation of our methodology. In case an application requires the presence of a collection of identical components interacting with each other, e.g. the existence of two molecules of hydrogen in $H_2O$, this could be modelled via two autonomous tokens namely $H_1$ and $H_2$.  We expect that our theory can be extended to relax both of these assumptions by associating further information to tokens and transitions in a system in order to distinguish between threads of execution associated to different tokens of the same base as well multiple occurrences of the same transition.

\section{Reversing Petri Nets}\label{sec: Reversible Petri Nets}
 
We are now ready to define a model of reversing Petri nets as follows:

\begin{definition}{\rm
 A \emph{\PN}(RPN) is a tuple $(A, P,B,T,F)$ where:
\begin{enumerate}
\item $A$ is a finite set of \emph{bases} or \emph{tokens} ranged over by $a$, $b$,\ldots.  $\overline{A} = \{\overline{a}\mid a\in A\}$
contains a ``negative" instance for each token and we write ${\cal{A}}=A \cup \overline{A}$.
\item $P$ is a finite set of \emph{places}.
\item $B\subseteq A\times A$ is a set of \emph{bonds} ranged over by $\beta$, $\gamma$,\ldots.
We use the notation $a \bond b$ for a bond $(a,b)\in B$. $\overline{B} = \{\overline{\beta}\mid \beta\in B\}$
contains a ``negative" instance for each bond and we write ${\cal{B}}=B \cup \overline{B}$.
\item $T$ is a finite set of \emph{transitions}.
\item $F : (P\times T  \cup T \times P)\rightarrow 2^{{\cal{A}}\cup {\cal{B}}}$ is a set of directed \emph{arcs}.
\end{enumerate}
}\end{definition}

A \PN is built on the basis of a set of \emph{bases} or, simply, \emph{tokens}. We take an individual-token interpretation thus
considering each token to have a unique name. In this way, tokens may be distinguished from
each other, their persistence can be guaranteed and their history inferred from the structure
of a Petri net (as implemented by function $F$, discussed below). Tokens correspond to
the basic entities that occur in a system. 
They may occur as stand-alone elements but as computation proceeds they may also merge together to form \emph{bonds}. \emph{Places}
and \emph{transitions} have the standard meaning. 

Directed arcs connect places to transitions and vice
versa and are labelled by a subset of ${\cal{A}}\cup {\cal{B}}$ where  $\overline{A} = \{\overline{a}\mid a\in A\}$
contains the set of ``negative'' tokens expressing token absence and  $\overline{B} = \{\overline{\beta}\mid \beta\in B\}$ is a set of ``negative" bonds expressing bond absence.
For a label $\ell= F(x,t)$ or $\ell = F(t,x)$, we assume that each token $a$ 
can appear in $\ell$ at most once, either as $a$ or as $\overline{a}$,
and that if a bond $(a,b)\in\ell$ then $a,b\in\ell$. Furthermore, for $\ell = F(t,x)$,
it must be that $\ell\cap (\overline{A}\cup \overline{B}) = \emptyset$, that is, negative tokens/bonds may only occur
on arcs incoming to a transition. 
Intuitively, these labels express the requirements for a transition
to fire when placed on arcs incoming the transition, and the effects of the transition when placed on the
outgoing arcs. Thus, if $a\in F(x,t)$ this implies that token $a$ is required for the transition $t$ to
fire, and similarly for a bond $\beta\in F(x,t)$. On the other hand, $\overline{a}\in F(x,t)$, expresses that
token $a$ should not be present in the incoming places of $t$ for the transition to fire and similarly for bond $\overline\beta \in F(x,t)$. Note
that negative tokens/bonds are close in spirit to inhibitor arcs of extended Petri nets. Finally, note that 
$F(x,t)= \emptyset$ implies that there is no arc from place $x$ to transition $t$ and similarly for
$F(t,x) = \emptyset$.

We introduce the following notations. We write 
$\circ t =   \{x\in P\mid  F(x,t)\neq \emptyset\}$ and  
$ t\circ = \{x\in P\mid F(t,x)\neq \emptyset\}$
for the incoming and outgoing places of transition
$t$, respectively. Furthermore, we write
$\guard{t}  =   \bigcup_{x\in P} F(x,t)$ for the union of all labels on the incoming arcs of  transition $t$, and
$\effects{t}  =   \bigcup_{x\in P} F(t,x)$ for the union of all labels on the outgoing arcs of transition $t$.
\begin{definition}\label{well-formed}{\rm 
For a \PN to be \emph{well-formed}, it should satisfy the following conditions for all $t\in T$:
\begin{enumerate}
\item $A\cap \guard{t} = A\cap \effects{t}$,
\item If $ a \bond b \in \guard{t}$ then $ a \bond b \in \effects{t}$,
\item $ F(t,x)\cap F(t,y)=\emptyset$ for all $x,y\in P$, $x\neq y $. 
\end{enumerate}
}\end{definition}

According to the above we have that: (1) transitions do not
 erase tokens, (2) transitions do not destroy bonds, that is, if a bond $a\bond b$ exists in an input place of a transition, then it is
maintained in some output place, and 
(3) tokens/bonds cannot be cloned into more than one outgoing places.

In a graphical representation, tokens are indicated by $\bullet$, places by circles, transitions by boxes, and bonds by lines between tokens. As with standard Petri nets, we employ the notion of a \emph{marking}. A marking is a distribution
of tokens and bonds across places,  $M: P\rightarrow A\cup B$ where $a \bond b \in M(x)$, for some $x\in P$, implies
$a,b\in M(x)$. 
In addition, we employ the notion of a \emph{history} which assigns a memory to each transition,
$H : T\rightarrow \epsilon \cup\mathbb{N}$. Intuitively, a history of $\epsilon$ captures that the transition has not taken place and a history of $n\in\mathbb{N}$ captures that the transition
was executed and not reversed where $n$ indicates the order of execution amongst
non-reversed actions.
$H_0$ denotes the initial history where $H_0(t) = \epsilon$ for all $t\in T$. In a graphical representation 
histories are presented over the respective transitions as $[m]$, where
$m = H(t)$ for transition $t$.
A pair of a marking and history describes a \emph{state} of a PN based on which execution
is determined. We use the notation $\state{M}{H}$ to denote states.

As the last piece of our machinery, we define a notion that identifies connected
components of tokens within a place. 
Note that more than one connected component may arise
in a place due to the fact that various unconnected tokens may be moved to a place 
simultaneously by a transition, while the reversal of transitions which results
in the destruction of bonds may break down a connected component into various
subcomponents. We define $\connected(a,C)$, where $a$ is a base and $C\subseteq A\cup B$ a set of connections,
to be the tokens connected
to $a$ via bonds as well as the bonds creating these connections according to 
set $C$.

\[\connected(a,C)=(\{a\} \cap C)\cup\{\beta, b, c \mid \exists  w \mbox{ s.t. }  \paths(a,w,C), \beta\in w \mbox{, and } \beta=(b,c) \}\]
where $\paths(a,w,C)$ if $w=\langle\beta_1,\ldots,\beta_n\rangle$, and for all $1\leq i \leq n$, $\beta_i=(a_{i-1},a_i)\in C\cap B$, $a_i\in C\cap A$, and $a_0 = a$.
We also write $\connected(S,C)$, where $S\subseteq A$, for $\bigcup_{a\in S}\connected(a,C)$.

Returning to the example of Figure~\ref{catalyst}, we may see
a reversing net with three tokens $a$, $b$, and $c$, transition $t_1$, which bonds tokens $a$ and $c$ within place $x$,
and  transition $t_2$, which bonds the $a$ of bond $a\bond c$ with token $b$ into place $y$.
The final marking after executing $t_1$ and consecutively $t_2$ assigns a complex of three bonded tokens named $c\bond a\bond b$ into place $y$. By applying out-of-causal order reversibility we are able to reverse transition $t_1$ and release token $c$ back to its initial place which is place $u$. The current marking also indicates the bond $a\bond b$ remaining intact in place $y$.

\section{Semantics}\label{sec:Methodology}
We may now define the various types of execution within reversing Petri nets. In what follows we
restrict our attention to PNs $(A,P,B,T,F)$ where $F$ defines an acyclic graph with initial
marking $M_0$ such that for all $a\in A$, $|\{ x \mid a\in M_0(x)\} | = 1$.  

\subsection{Forward Execution}

\begin{definition}\label{forward}{\rm
Consider a \PN $(A, P,B,T,F)$, a transition $t\in T$, and a state $\state{M}{H}$. We say that
 $t$ is \emph{forward enabled} in $\state{M}{H}$ if the following hold:
 \begin{enumerate}
 \item  if $a\in F(x,t)$,  for some $x\in\circ t$, then $a\in M(x)$, and if   
$\overline{a}\in F(x,t)$
  for some $x\in\circ t$, then $a\not\in M(x)$, 
   \item  if $\beta\in F(x,t)$,  for some $x\in\circ t$, then $\beta\in M(x)$, and if   
$\overline{\beta}\in F(x,t)$
  for some $x\in\circ t$, then $\beta\not\in M(x)$, 
  \item if $a\in F(t,y_1)$ and  $b\in F(t,y_2)$ where $y_1 \neq y_2$ then $b \not\in \connected(a,M(x))$ where $x \in \circ t$, and 
\item if $\beta\in F(t,x)$ for some $x\in t\circ$ and $\beta\in M(y)$ for some $y\in \circ t$ then $\beta\in F(y,t)$. 
 \end{enumerate}
}\end{definition}

Thus, $t$ is enabled in state $\state{M}{H}$ if (1), (2)  all tokens and bonds required for the transition to take place
are available in the incoming places of $t$ and  none of the tokens/bonds whose absence is
required exists in an incoming place of the transition, (3) if a transition forks into out-places $y_1$ and
$y_2$ then the tokens transferred to these places are not connected to each other in the
incoming places to the transition,  and (4) if a pre-existing bond 
appears in an outgoing arc of a transition, then it is also a precondition 
of the transition to fire. 
Contrariwise, if the bond appears in an outgoing arc of a transition ($\beta\in F(t,x)$ for some $x\in t\circ$)
but is not a requirement for the transition to fire ($\beta\not\in F(y,t)$ for all $y\in \circ t$),
then  the bond should not be present in an in-place of the transition ($\beta\not\in M(y)$ for all $y\in \circ t$). 
Thus, we define the effect of a transition as 
\[\effect{t} = \effects{t} - \guard{t}\]

We observe that the effect of a transition is 
the set of new bonds created by the transition  since,
by Definition~\ref{forward}(4),  the bonds that are created by the transition are
exactly those that occur in the postcondition of a transition but not in its precondition.  
This will subsequently enable the enunciation of transition reversal by the destruction of exactly
the bonds in $\effect{t}$.

\begin{definition}{\rm \label{forw}
Given a \PN $(A, P,B,T,F)$, a state $\langle M, H\rangle$, and a transition $t$ enabled in $\state{M}{H}$, we write $\state{M}{H}
\trans{t} \state{M'}{H'}$
where:
\[
\begin{array}{rcl}
	M'(x) & = & \left\{
	\begin{array}{ll}
		M(x)-\bigcup_{a\in F(x,t)}\connected(a,M(x)),  & \textrm{if } x\in \circ{t} \\
		M(x)\cup F(t,x)\cup \bigcup_{ a\in F(t,x), y\in\circ{t}}\connected(a,M(y)), & \textrm{if }  x\in t\circ\\
        	M(x), & \textrm{otherwise}
	\end{array}
	\right.
\end{array}
\]
and
\[
\begin{array}{rcl}
	H'(t') & = & \left\{
	\begin{array}{ll}
		\max\{k| k = H(t''), t''\in T\} +1,\hspace{0.2in} & \textrm{if } t' = t \\
        	H(t'), & \textrm{ otherwise}
	\end{array}
	\right.
\end{array}
\]
}\end{definition}

According to the definition, when a transition $t$ is executed, all tokens and bonds occurring in its incomings arcs are 
relocated from the input places to the output places along with their connected components.
Moreover, history function $H$ is extended to $H'$ by assigning to transition $t$ the next available integer key.

\begin{figure}[t]
\centering
\subfigure{\includegraphics[width=5cm]{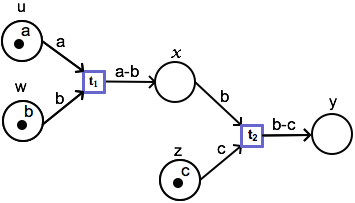}}
\subfigure{\includegraphics[width=.6cm]{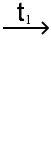}}
\subfigure{\includegraphics[width=5cm]{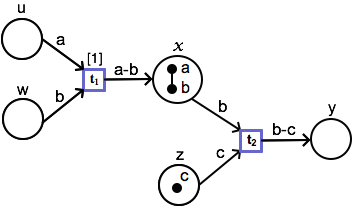}}
\subfigure{\includegraphics[width=.6cm]{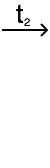}}
\subfigure{\includegraphics[width=5cm]{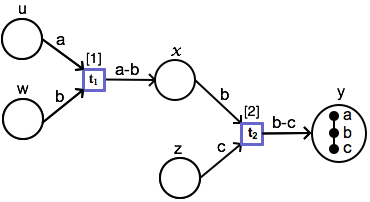}}
\subfigure{\includegraphics[width=.6cm]{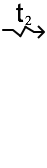}}
\subfigure{\includegraphics[width=5cm]{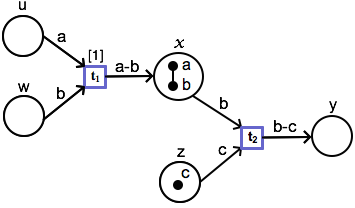}}
\subfigure{\includegraphics[width=.6cm]{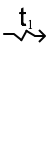}}
\subfigure{\includegraphics[width=5cm]{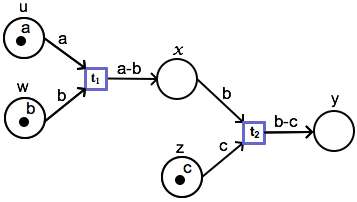}}
\vspace{-1em}
\caption{Forward and backtracking execution}
\label{b-example}
\end{figure}

An example of forward transitions can be seen in the first three steps of
Figure~\ref{b-example} where transitions $t_1$ and $t_2$ take place with the histories of the two transitions 
becoming $[1]$ and $[2]$, respectively. Note that to avoid overloading figures, we omit writing the bases of
bonds on the arcs of an \RPN and recall that within places we indicate bases by $\bullet$ and bonds by lines between 
relevant bases.

We may prove the following result:
\begin{proposition} \label{prop1} [Token and bond preservation.]    {\rm Consider a \RPN $(A, P,B,T,F)$, a state $\langle M, H\rangle$ such that
for all $a\in A$,  $|\{x\in P\mid a\in M(x)\}| = 1$, and a transition 
$\state{M}{H} \trans{t} \state{M'}{H'}$. Then (1)  for each base $a\in A$,  $|\{x\in P \mid a\in M'(x)\}|=1$,
and (2) for each bond $\beta\in B$, $|\{x\in P \mid \beta\in M(x)\}| \leq |\{x\in P \mid \beta\in M'(x)\}|\leq 1$.
}\end{proposition}
\paragraph{Proof:}
The proposition verifies that bases are preserved during forward execution in the sense that transitions neither erase nor clone them. As far as bonds are concerned, the proposition states that forward execution may only
create new bonds.
The proof of the result follows the definition of forward execution and relies on the well-formedness of
reversing Petri nets. 

Consider a \RPN $(A, P,B,T,F)$, a state $\langle M, H\rangle$  such that $|\{x\in P\mid a\in M(x)\}| = 1$ for
all $a\in A$, and suppose 
$\state{M}{H} \trans{t} \state{M'}{H'}$.  

For the proof of clause (1) let $a\in A$. Two cases exist:
\begin{enumerate}
\item $a\in \connected(b,M(x))$ for some $b\in F(x,t)$. Note that $x$ is unique by the
assumption that $|\{x\in P\mid a\in M(x)\}| = 1$. Furthermore, according to Definition~\ref{forw}, 
we have that $M'(x) = M(x) - \{\connected(c,M(x)) \mid c\in F(x,t)\}$, which implies that $a\not\in M'(x)$.
On the other hand, note that by  Definition~\ref{well-formed}(1),
$b\in \effects{t}$. Thus, there exists $y\in t\circ$, such that $b\in F(t,y)$. Note that this $y$
is unique by Definition~\ref{well-formed}(3).  As a result, by Definition~\ref{forw}, 
$M'(y) = M(y)\cup F(t,y)\cup \{\connected(c,M(y')) \mid c\in F(t,y), y'\in\circ{t}\}$. Since
$b\in F(x,t)\cap F(t,y)$, $a\in \connected(b,M(x))$, this implies that $a\in M'(y)$. 

Now suppose that $a\in \connected(c,M(x))$ for some  $c\neq b$, $c\in F(t,y')$. Then, by Definition~\ref{forward}(3), 
it must be that $y = y'$. As a result, we have that $\{z\in P\mid a\in M'(z)\} = \{y\}$ and 
the result follows.
\item $a\not\in \connected(b,M(x))$ for all $b\in F(x,t)$, $x\in P$. This implies that 
$\{x\in P\mid a\in M'(x)\} = \{x\in P\mid a\in M(x)\}$ and the result follows.
\end{enumerate}

To prove clause (2) of the proposition, consider a bond
$\beta\in B$, $\beta=(a,b)$. We observe that, since $|\{x\in P\mid a\in M(x)\}| = 1$ for
all $a\in A$, $|\{x\in P\mid \beta\in M(x)\}| \leq 1$. The proof follows by case analysis as
follows:
\begin{enumerate}
\item Suppose $|\{x\in P\mid \beta\in M(x)\}| =0$. Two cases exist:
	\begin{itemize}
	\item Suppose $\beta\not\in F(t,x)$ for all $x\in P$. Then, by Definition~\ref{forw}, $\beta\not\in M'(x)$ 
	for all $x\in P$. Consequently, $|\{x\in P\mid \beta\in M'(x)\}| =0$ and the result follows.
	\item Suppose $\beta\in F(t,x)$ for some $x\in P$. Then, by  Definition~\ref{well-formed}(3),
	$x$ is unique, and by Definition~\ref{forw}, $\beta\in M'(x)$.
	Consequently, $|\{x\in P\mid \beta\in M'(x)\}| =1$ and the result follows.
	\end{itemize}
\item Suppose $|\{x\in P\mid \beta\in M(x)\}| =1$. Two cases exist:
	\begin{itemize}
	\item $\beta\not\in \connected(c,M(x))$ for all $c\in F(x,t)$. 
	This implies that $\{x\in P\mid \beta\in M'(x)\} = \{x\in P\mid \beta\in M(x)\}$ and the result follows.
	\item $\beta\in \connected(c,M(x))$ for some $c\in F(x,t)$. Then, according to Definition~\ref{forw}, 
	we have that $M'(x) = M(x) - \{\connected(c,M(x)) \mid c\in F(x,t)\}$, which implies that $\beta\not\in M'(x)$.
	On the other hand, note that by the definition of well-formedness, Definition~\ref{well-formed}(1),
	$c\in \effects{t}$. Thus, there exists $y\in t\circ$, such that $c\in F(t,y)$. Note that this $y$
	is unique by Definition~\ref{well-formed}(3).  As a result, by Definition~\ref{forw}, 
	$M'(y) = M(y)\cup F(t,y)\cup \{\connected(c,M(y')) \mid c\in F(t,y), y'\in\circ{t}\}$. Since
	$c\in F(x,t)\cap F(t,y)$, $\beta\in \connected(c,M(x))$, this implies that $\beta\in M'(y)$. 

	Now suppose that $\beta\in \connected(d,M(x))$ for some $d\neq c$, and $c\in F(d,y')$. Then, by Definition~\ref{forward}, 
	and since $\connected(c,M(x)) = \connected(d,M(x))$,
	it must be that $y = y'$. As a result, we have that $\{z\in P\mid \beta\in M'(z)\} =  \{y\}$ and 
	the result follows.
	\end{itemize}
\proofend
\end{enumerate}

\subsection{Backtracking}
Let us now proceed to the simplest form of reversibility, namely, backtracking.
\begin{definition}{\rm

Consider a \PN $(A, P,B,T,F)$ a state $\state{M}{H}$ and a transition $t\in T$. We say that $t$ is \emph{$bt$-enabled} in
$\state{M}{H}$ if
$H(t)=k \in \mathbb{N}$ with $k\geq k'$ for all $k' \in \mathbb{N}$, $k'= H(t')$, $t'\in T$.
}\end{definition}

Thus, a transition $t$ is $bt$-enabled if 
it is the last transition to have been executed,
i.e., it has the highest $H$ value. 
The effect of reversing a transition in a Petri net in a backtracking fashion is as follows:

\begin{definition}\label{br-def}{\rm
Given a \PN $(A, P,B,T,F)$, a state $\langle M, H\rangle$, and a transition $t$ $bt$-enabled in $\state{M}{H}$, we write $ \state{M}{H}
\btrans{t} \state{M'}{H'}$
where:
\[
\begin{array}{rcl}
	M'(x) & = & \left\{
	\begin{array}{ll}
		M(x)\cup\bigcup_{ y \in t\circ, a\in F(x,t)\cap F(t,y)}\connected(a,M(y)-\effect{t}), \hspace{0.3in} & \textrm{if } x\in \circ{t} \\
		M(x)- \bigcup_{a\in F(t,x)}\connected(a,M(x)) , & \textrm{if }  x\in t\circ\\
        M(x), & \textrm{ otherwise}
	\end{array}
	\right.
\end{array}
\]
and
\[
\begin{array}{rcl}
	H'(t') & = & \left\{
	\begin{array}{ll}
		\epsilon, \hspace{0.2in}& \textrm{if } t' = t \\ 
        	H(t), & \textrm{ otherwise}
	\end{array}
	\right.
\end{array}
\]
}\end{definition}

Thus, when a transition $t$ is reversed in a backtracking fashion  all tokens and bonds in the
postcondition of the transition, as well as their connected components, 
will be transferred to the incoming places of the transition and any newly-created bonds will be broken.
Moreover, history function $H$ is refined to $H'$ by setting $H'(t)=\epsilon$, capturing that the specific transition has been reversed.
In the last two steps of Figure~\ref{b-example}  we observe transitions $t_2$ and $t_1$ being reversed  with the histories of the two transitions being eliminated.

We may prove the following result:
\begin{proposition}\label{prop2} [Token preservation and bond destruction.]  {\rm Consider a \RPN $(A, P,B,T,F)$, a state $\langle M, H\rangle$
such that for all $a\in A$, $|\{x\in P \mid a\in M(x)\}| = 1$, and a transition 
$\state{M}{H} \btrans{t} \state{M'}{H'}$. Then, (1) for each base $a\in A$,  
$|\{x\in P \mid a\in M'(x)\}| = 1$, and (2)
for each bond $\beta\in B$, $1 \geq |\{x \in P \mid \beta\in M(x)\}| \geq |\{x \in P \mid \beta\in M'(x)\}|$.
}\end{proposition}
\paragraph{Proof:}
The proposition verifies that bases are preserved during backtracking execution in the sense that
there exists exactly one instance of each base and backtracking transitions neither erase nor clone them. 
As far as bonds are concerned, the proposition states that at any time there may exist at most one 
instance of a bond and that backtracking transitions may only destroy bonds. 
The proof of the result follows the definition of backward execution and relies on the well-formedness of
reversing Petri nets. 
Consider a \RPN $(A, P,B,T,F)$, a state $\langle M, H\rangle$ such that $|\{x\in P \mid a\in M(x)\}| =1$
for all $a\in A$, and suppose  $\state{M}{H} \btrans{t} \state{M'}{H'}$.  

We begin with the proof of clause (1) and let $a\in A$. Two cases exist:
\begin{enumerate}

\item $a\in \connected(b,M(x))$ for some $b\in F(t,x)$. Note that by the assumption 
of $|\{x\in P \mid a\in M(x)\}| =1$, $x$ must be unique. 
Let us choose $b$ such that, additionally, $a\in \connected(b,M(x) - \effect{t})$. Note
that such a $b$ must exist, otherwise 
the forward execution of $t$ would not have transferred $a$ along with $b$ to place $x$.

According to Definition~\ref{br-def}, 
we have that $M'(x) = M(x) - \{\connected(c,M(x)) \mid c\in F(t,x)\}$, which implies that $a\not\in M'(x)$.
On the other hand, note that by the definition of well-formedness, Definition~\ref{well-formed}(1),
$b\in \guard{t}$. Thus, there exists $y\in \circ t$, such that $b\in F(y,t)$. Note that this $y$
is unique. If not, then there exist
$y$ and $y'$ such that $y\neq y'$ with $b\in F(y,t)$ and $b\in F(y',t)$.  By the assumption, however,
that there exists at most one token of each base, and Proposition~\ref{prop1}, $t$ would never be enabled, 
which leads to a contradiction.   As a result, by Definition~\ref{br-def}, 
$M'(y) = M(y)\cup\{\connected(c,M(y')-\effect{t}) \mid c\in F(y,t)\cap F(t,y')\}$. Since
$b\in F(y,t)\cap F(t,x)$, $a\in \connected(b,M(x)-\effect{t})$, this implies that $a\in M'(y)$. 

Now suppose that $a\in \connected(c,M(x)-\effect{t})$, $c\neq b$, and $c\in F(y',t)$. Since
 $a\in \connected(b,M(x) - \effect{t})$, it must be that
$\connected(b,M(x) - \effect{t})=\connected(c,M(x)-\effect{t})$. Since $b$ and $c$ are
connected to each other but the connection was not created by transition $t$ (the connection is
present in $M(x)-\effect{t}$), it must be the connection was already present before
the forward execution of $t$ and, by token uniqueness, we conclude that  $y=y'$.
\item $a\not\in \connected(b,M(x))$ for all $b\in F(t,x)$, $x\in P$. This implies that 
$\{x\in P\mid a\in M'(x)\} = \{x\in P\mid a\in M(x)\}$ and the result follows.
\end{enumerate}

Let us now prove clause (2) of the proposition. Consider a bond
$\beta\in B$, $\beta=(a,b)$. We observe that, since $|\{x\in P\mid a\in M(x)\}| = 1$ for
all $a\in A$, $|\{x\in P\mid \beta\in M(x)\}| \leq 1$. The proof follows by case analysis as
follows:
\begin{enumerate}
\item $\beta\in \connected(c,M(x))$ for some $c\in F(t,x)$, $x\in P$. By the assumption 
of $|\{x\in P \mid \beta\in M(x)\}| =1$, $x$ must be unique. Then, according to Definition~\ref{br-def}, 
we have that $M'(x) = M(x) - \{\connected(c,M(x)) \mid c\in F(x,t)\}$, which implies that $\beta\not\in M'(x)$.
Two cases exist: 
\begin{itemize}
\item If $\beta\in \effect{t}$, then $\beta\not\in M'(y)$ for all places $y\in P$.
\item If $\beta\not\in \effect{t}$ then let us choose $c$ such that 
$\beta\in \connected(c,M(x) - \effect{t})$. Note
that such a $c$ must exist, otherwise 
the forward execution of $t$ would not have connected $\beta$ with $c$. 
By the definition of well-formedness, Definition~\ref{well-formed}(1),
$c\in \guard{t}$. Thus, there exists $y\in \circ t$, such that $c\in F(y,t)$. Note that this $y$
is unique (if not, $t$ would not have been enabled).  As a result, by Definition~\ref{br-def}, 
$\beta \in M'(y)$.

Now suppose that $\beta\in \connected(d,M(x)-\effect{t})$, $d\neq c$, and $d\in M'(y')$. Since
 $\beta\in \connected(c,M(x) - \effect{t})$, it must be that
$\connected(c,M(x) - \effect{t})=\connected(d,M(x)-\effect{t})$.  Since $c$ and $d$ are
connected to each other but the connection was not created by transition $t$ (the connection is
present in $M(x)-\effect{t}$), it must be the connection was already present before
the forward execution of $t$ and, by token uniqueness, we conclude that  $y=y'$.
This implies that $\{z\in P\mid \beta\in M'(z)\} = \{y\}$. 
\end{itemize}
The above imply that $\{z\in P\mid \beta\in M(z)\} = \{x\}$ and  $\{z\in P\mid \beta\in M'(z)\} \subseteq \{y\}$ and 
the result follows.
\item $\beta\not\in \connected(c,M(x))$ for all $c\in F(t,x)$, $x\in P$. This implies that 
$\{x\in P\mid \beta\in M'(x)\} = \{x\in P\mid \beta\in M(x)\}$ and the result follows.
\proofend
\end{enumerate}

Let us now consider the combination of forward and backward moves in executions. 
We write $\fbtrans{}$ for $\trans{}\cup\btrans{}$.
The following result establishes that in an execution beginning in the initial state of a Petri net, bases are 
preserved, bonds can have at most one instance at any time and a new occurrence of  a bond may be created 
during a forward  transition that features the bond as its effect whereas a bond can be destroyed 
during the backtracking of a transition that features the bond as its effect. This last point clarifies
that the effect of a transition characterises the bonds that are newly-created during the transition's forward execution
and the ones that are being destroyed during its reversal.

\begin{proposition}\label{Prop}{\rm Given a \RPN $(A, P,B,T,F)$, an initial state 
$\langle M_0, H_0\rangle$ and an execution
$\state{M_0}{H_0} \fbtrans{t_1}\state{M_1}{H_1} \fbtrans{t_2}\ldots \fbtrans{t_n}\state{M_n}{H_n}$, the following hold:
\begin{enumerate}
\item For all $a\in A$  and $i$, $0\leq i \leq n$,   $|\{x\in P \mid a\in M_i (x)\}| = 1$.
\item For all $\beta \in B$ and $i$, $0\leq i \leq n$, $0 \leq |\{x \in P \mid \beta\in M_i(x)\}| \leq 1$, and,
\begin{enumerate}
\item if $t_i$ is a forward transition with $\beta\in \effect{t_i}$, then 
$\beta\in M_{i}(x)$ for some $x\in P$ and  $\beta\not\in M_{i-1}(y)$ for all $y\in P$,
\item if $t_i$ is a backtracking transition with $\beta\in \effect{t_i}$ then 
$\beta\in M_{i-1}(x)$ for some $x\in P$ and  $\beta\not\in M_{i}(y)$ for all $y\in P$, and
\item if $\beta\not\in \effect{t_i}$ then $\beta\in M_{i-1}(x)$ if and only if $\beta\in M_i(x)$.
\end{enumerate}
\end{enumerate}}
\end{proposition}
\paragraph{Proof:}
The proposition verifies that (1) tokens are preserved throughout the execution of an RPN, (2) bonds can be created (during forward execution), destructed (during backward execution),
or preserved through actions that do not operate directly on the bond.

To begin with, we observe that 
the proofs of clauses (1) and (2) follow directly from clauses (1) and (2) of Propositions~\ref{prop1} 
and~\ref{prop2}, respectively. Clause (3) follows from Definition~\ref{forward}(4) and, finally, clause (4) 
stems from Definition~\ref{br-def} and can be proved as in the proof of Proposition~\ref{prop2}(2), case (1).
\proofend

In this setting we may establish a loop lemma:
\begin{lemma}[Loop]\label{loopb}{\rm 
For any forward transition $\state{M}{H}\trans{t}\state{M'}{H'}$ there exists a backward transition
$\state{M'}{H'} \btrans{t} \state{M}{H}$ and vice versa. 
}\end{lemma}
\paragraph{Proof:}
Suppose $\state{M}{H}\trans{t}\state{M'}{H'}$. Then $t$ is clearly $bt$-enabled in $H'$. Furthermore,
$\state{M'}{H'} \btrans{t} \state{M''}{H''}$ where $H'' = H$. In addition, all tokens and bonds
involved in transition $t$ (except those in $\effect{t}$) will be returned from the out-places
of transition $t$ back to its in-places. Specifically, for all $a\in A$, it
is easy to see by the definition of $\btrans{}$  that $a\in M''(x)$ if and only if $a\in M(x)$.
Similarly, by Proposition~\ref{Prop}, for all $\beta\in B$,  $\beta\in M''(x)$ if and only if
$\beta\in M(x)$. The opposite direction can be argued similarly, only this
time tokens and bonds involved in transition $t$ will be moved from the in-places to
the out-places of transition $t$.
\proofend

\subsection{Causal Reversing}

We now move on to consider causality between transitions in a Petri net and reversibility in a causal-respecting order. The following definition shows that given a transition if all the causally linked transitions have either been reversed or not executed then the transition is co-enabled. 
Thus, we define the notion of causally-ordered enabledness as follows:

\begin{definition}\label{co-enabled}{\rm
Consider a \PN $(A, P,B,T,F)$, a state $\state{M}{H}$, and a transition $t\in T$. Then $t$ is
$co$-enabled in  $\state{M}{H}$ if
$H(t)\in \mathbb{N}$ and, for all $a\in F(t,x)$, if $a\in M(y)$ for some $y$ and $\connected(a,M(y))\cap \guard{t'} \neq \emptyset$
for some $t'$  then either $H(t')=\epsilon$ or $H(t')\leq H(t)$. 
}\end{definition}
We may prove the following equivalent enunciation of the definition which reduces $co$-enabledness to
the availability of tokens in the outplaces of an executed transition. Acyclicity of RPNs is central for the proof of this result.


\begin{proposition}\label{prop4}{\rm 

Consider a \PN $(A, P,B,T,F)$, a state $\state{M}{H}$, and a transition $t\in T$. Then $t$ is
$co$-enabled in  $\state{M}{H}$ if and only if $H(t)\in \mathbb{N}$ and
for all $a,\beta\in F(t,x)$ we have $a,\beta\in M(x)$.
}\end{proposition}
\paragraph{Proof:}
Let us consider the ``if'' direction. This can be proved by contradiction. Suppose that transition $t$ is 
\emph{co}-enabled in $\state{M}{H}$ 
and there exist bonds and/or tokens that are not located on its out-places but are required on its out-going 
arcs for the transition to reverse, i.e., there exist $a$ and/or $\beta \in F(t,x)$ but $a,\beta \not\in M(x)$.  
Since $H(t)\in\mathbb{N}$, the transition has been executed and not reversed. As a consequence of
Definition~\ref{forw} it must be that all bonds and tokens in $\effects{t}$ have been moved to the out-places
of $t$. The fact, however, that some of these are no longer available in these out-places suggests that
some transition $t'$ with $t\circ\cap\circ t'=\emptyset$ has been consequently executed and further moved these tokens in
its out-places. This implies that either $a\in \guard{t'}$, or there exists $b\in \guard{t'}$ with $a$ being connected 
to $b$ when $t'$ was executed, thus moving $a$ into the out-places of $t'$ when $t'$ was executed.
Note that, by our assumption of acyclicity of RPNs, this out-places must be different to the out-places of $t$.
However, by the conditions of \emph{co}-enabledness, it must be that $H(t')=\epsilon$. This results in a contradiction
and the result follows.

For the ``only if'' direction, suppose that for all $a,\beta\in F(t,x)$, $a,\beta\in M(x)$. Then, clearly, for 
all $t'$, $t\circ\cap \circ t'=\emptyset$, if $\connected(a,M(y))\cap \guard{t'}\neq \emptyset$ it must be that
$H(t')=\epsilon$ for, if not, then $a$ would have been moved to the out-places of $t'$ and, by
the acyclicity of the Petri net, $a\not \in M(x)$. Using an inductive argument, this is true for all
transitions $t'$ with $\connected(a,M(y))\cap \guard{t'}\neq \emptyset$. This completes the proof.
\proofend

Reversing a transition in a causally-respecting order is implemented in exactly the same way as in
backtracking, i.e., the tokens are moved from the out-places to the in-places of the transition, all bonds
created by the transition are broken, and the reversal eliminates the history function.

\begin{definition}{\rm
Given a \PN $(A, P,B,T,F)$, a state $\langle M, H\rangle$, and a transition $t$ $co$-enabled in $\state{M}{H}$ we write $\state{M}{H}
\ctrans{t} \state{M'}{H'}$ for $M'$ and $H'$ as in Definition~\ref{br-def}.
}\end{definition}

\begin{figure}[t]
\centering
\hspace{1.5em}
\subfigure{\includegraphics[width=.55cm]{arrow1.png}}
\subfigure{\includegraphics[width=.55cm]{arrow2.png}}
\subfigure{\includegraphics[width=.55cm]{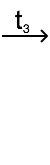}}
\subfigure{\includegraphics[width=5.4cm]{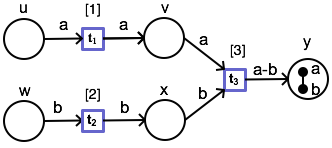}}
\subfigure{\includegraphics[width=.55cm]{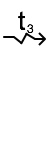}}
\subfigure{\includegraphics[width=5.4cm]{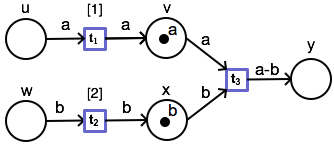}}
\hspace{.05em}
\subfigure{\includegraphics[width=.55cm]{arrow1r.png}}
\subfigure{\includegraphics[width=5.4cm]{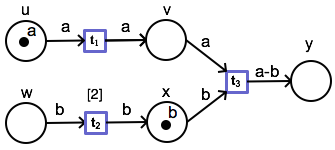}}
\subfigure{\includegraphics[width=.55cm]{arrow2r.png}}
\subfigure{\includegraphics[width=5.4cm]{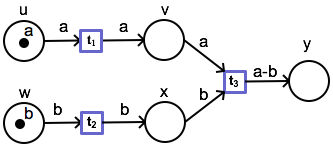}}
\caption{Causal Order example}\label{co-example}
\end{figure}

An example of causal-order reversibility can be seen in Figure~\ref{co-example}. Here we have two independent transitions, $t_1$ and $t_2$ causally preceding transition $t_3$. Assuming that transitions were executed in the order $t_1$, $t_2$,  $t_3$, the example demonstrates a causally-ordered reversal where  $t_3$ is (the only transition that can be) reversed, followed by the reversal of its two causes $t_1$ and $t_2$.  In general these can be reversed in any order although in the example $t_1$ is reversed before $t_2$. 

We may now establish the causal consistency of our semantics, as in~\cite{RCCS}. First we define 
some auxiliary notions. In what follows we write $\fctrans{}$ 
for $\trans{}\cup\ctrans{}$.

Given a transition $\state{M}{H}\fctrans{t}\state{M'}{H'}$, we say that the \emph{action} of the transition is
$t$ if $\state{M}{H}\trans{t}\state{M'}{H'}$ and $\underline{t}$ if $\state{M}{H}\ctrans{t}\state{M'}{H'}$
and we may write $\state{M}{H}\fctrans{\underline{t}}\state{M'}{H'}$. We use $\alpha$ to
range over $\{t,\underline{t} \mid t\in T\}$ and write $\underline{\underline{\alpha}} = \alpha$. We extend this
notion to sequences of transitions and, given an execution $\state{M_0}{H_0}\fctrans{t_1}\ldots 
\fctrans{t_n}\state{M_n}{H_n}$, we say that the \emph{trace} of the execution is
$\sigma=\alpha_1;\alpha_2\ldots;\alpha_n$, where $\alpha_i$ is the action of transition 
$\state{M_{i-1}}{H_{i-1}}\fctrans{t_i}\state{M_i}{H_i}$,  and write $\state{M}{H}\fctrans{\sigma}\state{M_n}{H_n}$.
\begin{definition}{\rm
Two actions $\alpha_1$ and $\alpha_2$ are said to be \emph{concurrent} if whenever
$\state{M}{H}\fctrans{\alpha_1}\state{M_1}{H_1}$ and $\state{M}{H}\fctrans{\alpha_2}\state{M_2}{H_2}$
then $\state{M_1}{H_1}\fctrans{\alpha_2}\state{M'}{H'}$ and $\state{M_2}{H_2}\fctrans{\alpha_1}\state{M'}{H''}$.
}\end{definition}
Thus, two actions are concurrent if execution of the one does not preclude the other. 

\begin{definition}\label{co-executions}{\rm  \emph{Causal equivalence on traces}, denoted by $\asymp$, is the least
equivalence relation closed under composition of traces such that 
(i) if $\alpha_1$ and $\alpha_2$ are concurrent actions then $\alpha_1;\alpha_2\asymp \alpha_2;\alpha_1$, and
(ii) $\alpha ; \underline{\alpha} \asymp \epsilon$.
}\end{definition}

The first clause states that in two causally equivalent traces concurrent actions
may occur in any order and the second clause states that it is possible to ignore transitions that have occurred in both
the forward and the reverse order.

 We additionally define a notion of history equivalence (overloading operator $\asymp$), according to which
two histories $H$ and $H'$ are equivalent if and only
if they record the same executed transitions that have not been reversed:

\begin{definition}\label{eq}{\rm  \emph{History equivalence}, denoted by $\asymp$, is defined
such that $H\asymp H'$ holds whenever  $H(t) = k$ for some $k\in\mathbb{N}$ if and only if $H'(t) = m$ for some $m\in\mathbb{N}$.
We extend this notion to states and write $\state{M}{H}\asymp\state{M}{H'}$ if and only if $H\asymp H'$.
}\end{definition}

When it is clear from the context we drop the subscript in the above relations and simply write $\asymp$ for all $\asymp_e$, $\asymp_h$ and $\asymp_s$. We may now prove the following results.

\begin{proposition}{\rm Given a \RPN $(A, P,B,T,F)$, an initial state 
$\langle M_0, H_0\rangle$ and an execution
$\state{M_0}{H_0} \fctrans{t_1}\state{M_1}{H_1} \fctrans{t_2}\ldots \fctrans{t_n}\state{M_n}{H_n}$, the following hold:
\begin{enumerate}
\item For all $a\in A$  and $i$, $0\leq i \leq n$,   $|\{x\in P \mid a\in M_i (x)\}| = 1$.
\item For all $\beta \in B$ and $i$, $0\leq i \leq n$, $0 \leq |\{x \in P \mid \beta\in M_i(x)\}| \leq 1$, and,
\begin{enumerate}
\item if $t_i$ is a forward transition with $\beta\in \effect{t_i}$, then 
$\beta\in M_{i}(x)$ for some $x\in P$ and  $\beta\not\in M_{i-1}(y)$ for all $y\in P$,
\item if $t_i$ is a reversing transition with $\beta\in \effect{t_i}$ then 
$\beta\in M_{i-1}(x)$ for some $x\in P$ and  $\beta\not\in M_{i}(y)$ for all $y\in P$, and
\item if $\beta\not\in \effect{t_i}$ then $\beta\in M_{i-1}(x)$ if and only if $\beta\in M_i(x)$.
\end{enumerate}
\end{enumerate}}
\end{proposition}

\paragraph{Proof:} The proof follows along the same lines as that of Proposition~\ref{Prop} with $\btrans{}$ replaced
by $\ctrans{}$.
\proofend

\begin{lemma}[Loop]\label{loop}{\rm 
For any forward transition $\state{M}{H}\trans{t}\state{M'}{H'}$ there exists a backward transition
$\state{M'}{H'} \ctrans{t} \state{M}{H}$ and vice versa.
}\end{lemma}

\paragraph{Proof:} The proof follows along the same lines as that of Lemma~\ref{loopb} with $\btrans{}$
replaced by $\ctrans{}$.
\proofend

The main result, Theorem~\ref{main} below, states  that if we have two computations beginning in the same initial state, then they lead to equivalent
states if and only if the sequences of executed transitions of the two computations are causally equivalent. Specifically,
if two executions from the same state reach the same marking by executing transitions $\sigma_1$
and $\sigma_2$ containing the same executed but not reversed actions, then $\sigma_1$ and $\sigma_2$ are causally
equivalent and vice versa. This guarantees the consistency of the approach since reversing transitions is in a sense equivalent
to not executing the transitions in the first place. Reversal will not give rise
to previously unreachable states, on the contrary, it will give rise to exactly the same markings and causally equivalent
histories due to the different keys being possibly assigned due to the different ordering or (lack of) reversal of transitions.

\begin{theorem}\label{main}{\rm Consider traces $\sigma_1$ and $\sigma_2$. Then, $\sigma_1\asymp\sigma_2$ if any only if  $\state{M_0}{H_0} \fctrans{\sigma_1} \state{M}{H}$ and $\state{M_0}{H_0} \fctrans{\sigma_2} \state{M}{H'}$ where  $\state{M}{H}\asymp\state{M}{H'}$.
}
\end{theorem}

For the proof of Theorem~\ref{main} we employ the following intermediate results. The lemma below
states that causal equivalence allows the permutation of inverse transitions that have no causal relations between them. 
Therefore, computations are allowed to reach for the maximum freedom of choice going backward and then continue forward.

\begin{lemma}\label{perm}{\rm \
Let $\sigma$ be a trace. Then there exist traces $r,r'$ both forward such that  
$\sigma\asymp\underline{r};r'$.
}\end{lemma}
\vspace{-0.2in} 

\paragraph{Proof:}
We prove this by induction on the length of  $\sigma$ and the
distance from the beginning of $\sigma$ to the earliest pair of transitions
that contradicts the property $\underline{r};r'$. If there is no such
contradicting pair then the property is trivially satisfied.  If not, we distinguish the following cases:
\begin{enumerate}
\item If the first contradicting pair is of the form $t;\underline{t}$
then since $t;\underline{t} = \epsilon$, we may remove the two transitions from
the sequence. Thus, the length of $\sigma$ decreases and the proof follows
by induction.
\item If the first contradicting pair is of the form $t;\underline{t}'$ then
we observe that $t$ and $\underline{t}'$ must be concurrent transitions: Since
action $t'$ is being reversed it implies that all the actions that are causally dependent on it have either
not been executed up to this point or they have already been reversed. This implies that action $t'$
is not causally dependent on $t$ and it can be executed before or after its
execution. That is, action $t$ is forward enabled, meaning that tokens exist to its input places
irrespectively of whether $t'$ has or has not been reversed.
As a result $t$ and $\underline{t'}$ can be swapped, resulting in a
later earliest contradicting pair. Thus by induction the result follows.
\proofend
\end{enumerate}

From the above lemma we may conclude the following corollary establishing that causal-order
reversibility is consistent with standard forward execution in the sense that RPNs will not generate
states that are unreachable in forward execution:

\begin{corollary}{\rm\ \
Suppose that  $H_0$ is the initial history.
If $\state{M_0}{H_0} \fctrans{\sigma} \state{M}{H_1}$ and $\sigma$ is a
trace with both forward and backward transitions if and only if
there exists a transition $\state{M_0}{H_0}\fctrans{\sigma'}\state{M}{H_2}$ 
and $\sigma'$ a trace of forward transitions.
 }\end{corollary}
%
\vspace{-0.25in}
\paragraph{Proof:} Proving the ``only if' part of the corollary is trivial since every forward computation in RPNs can
be simulated in reversible RPNs by only moving forward. To prove the ``if" part, according to
Lemma~\ref{perm}, $\sigma\asymp \underline{r};r'$ where both $r$ and $r'$ are forward
traces. Since, however, $H_0$ is the initial history it must be that $r$ is empty. This
implies that $\state{M}{H_0}\fctrans{r'}\state{M'}{H_2}$, $H_1\asymp H_2$ and $r'$ is a 
forward trace. Consequently, writing $\sigma'$ for $r'$, the result follows.
\proofend

\begin{lemma}\label{short}{\rm\ \
Suppose $\state{M}{H}\fctrans{\sigma_1}\state{M'}{H_1}$ and
$\state{M}{H}\fctrans{\sigma_2}\state{M'}{H_2}$, where $H_1\asymp H_2$ and
$\sigma_2$ is a forward trace. Then, there exists a forward trace
$\sigma_1'$ such that $\sigma_1 \asymp \sigma_1'$.
}\end{lemma}

 \vspace{-0.25in}
\paragraph{Proof:}
If  $\sigma_1$ is forward then $\sigma_1 = \sigma_1'$ and the result follows
trivially. Otherwise, we may prove the lemma by induction on the length of
$\sigma_1$.
We begin by noting that, by Lemma~\ref{perm},
$\sigma_1\asymp\underline{r};r'$. Let $\underline{t};t'$ be
the two successive transitions in $\underline{r};r'$  with opposite directions.
Given that $\sigma_2$ is a forward transition that simulates $\sigma_1$, it must
be that $r'$ contains a forward execution of transition $t$. If not, $H_2$ will
contain a forward token for transition $t$ and $H_1$ a reverse token leading
to a contradiction since, by definition, it would not hold that $H_1\asymp H_2$.
Consider the earliest 
occurrence of $t$ in $r'$ and let $t*$ be any transition between the reversal of $t$ and
this forward occurrence. Since $t*\in r'$, $t*$ is a forward transition that is not
reversed in the computation. Consequently, $t*$ must also occur in $\sigma_2$.
As a result, we observe that $t*$ is enabled both after a forward execution on
$t$, as in $\sigma_2$, as well as after a backward execution of $t$, as in $\underline{r},r'$.
Thus, it must be that $t$ and $t*$ are concurrent transitions. Since this
holds for any $t*$ between the occurrences of $\underline{t}$ and $t$, this implies
that we may permute $t$ with each such $t*$ 
within the trace to yield the sequence $\underline{t};t$. Since
$\underline{t};t \asymp \epsilon$, we may remove the pair of opposite
transitions and obtain a shorter equivalent trace, also
equivalent to $\sigma_2$ and conclude by induction.
\proofend

We may now proceed with the proof of Theorem~\ref{main}:

\paragraph{Proof of Theorem~\ref{main}:}
Suppose $\state{M_0}{H_0} \fctrans{\sigma_1} \state{M}{H}$ and $\state{M_0}{H_0} \fctrans{\sigma_2} \state{M}{H'}$
with  $\state{M}{H}\asymp\state{M'}{H'}$. 
We prove that $\sigma_1\asymp \sigma_2$ by using a lexicographic induction on a pair consisting 
of the sum of the lengths of $\sigma_1$ and $\sigma_2$ and the depth of the earliest disagreement 
between them. By Lemma~\ref{perm} we may suppose that $\sigma_1$ and $\sigma_2$
 are permutable to coincide with the property
 $\underline{r};r'$. Call $t_1$ and $t_2$ the earliest transitions where they disagree. There are three
  main cases in the argument depending on whether these are forward or backward.
\begin{enumerate}

\item If $t_1$ is backward and $t_2$ is  forward, we have $\sigma_1=\underline{r};t_1;u$ 
and $\sigma_2=\underline{r};t_2;v$ for some $r,u,v$. Lemma~\ref{short} applies to $t_2;v$ 
which is forward and $t_1;u$ which contains both forward and backward actions and thus,
by the lemma, it  has a shorter forward equivalent. Thus, $\sigma_1$ has a shorter forward 
equivalent and the result follows by induction.

\item If $t_1$ and $t_2$ are both forward  then it must be that case that $\sigma_1 = r;t_1; u$
and $\sigma_2 = r; t_2, v$, for some $r$, $u$, $v$, where $t_1\in v$ and $t_2\in u$. This
is the case since, if we consider the final states of the executions $H\asymp H'$, holds. 
Consequently, since $H(t_1)\in \mathbb{N}$, $H_2(t_1)\in\mathbb{N}$ should also hold.
In words, $t_1$ must also be executed by $\sigma_2$ and similarly for $t_2$ and $\sigma_1$.
The question that now arises is whether $t_1$ and $t_2$ are concurrent. Applying the same
arguments as in the proof of Lemma~\ref{short} we may conclude that the occurrence of
$t_1$ in $\sigma_2$ can be moved forward within $u$, yielding $\sigma_2 \asymp r;t_1;u'$.
This results in an equivalent execution of the same length with a later earliest divergence than with $\sigma_2$ 
and the result follows by the induction hypothesis.

\item If $t_1$ and $t_2$ are both backward, we have $\sigma_1=\underline{r};t_1;u$ 
and $\sigma_2=\underline{r};t_2;v$ for some $r,u,v$. Suppose that both $u$ and $v$
are minimized according to Lemma~\ref{short}. Then it must be that both $t_1$ and
$t_2$ are executed in $v$ and $u$ respectively. We may then argue that, e.g. $t_1$
can be moved forward within $v$ yielding $\sigma_2\asymp\underline{r};t_1;v'$
This results in an equivalent execution of the same length with a later earliest divergence than with $\sigma_2$ 
and the result follows by the induction hypothesis.
\proofend
\end{enumerate}

\subsection{Out-of-causal-order Reversibility}
Finally, we consider out-of-causal-order reversibility. Let us begin by considering the example
of Figure~\ref{o-example}. In the first net shown in the figure, we see that transitions $t_1$, $t_2$,
and $t_3$ have been executed in this order and now all tokens are in the final place $z$.
Suppose that transition $t_1$ is reversed out of order. As we have already discussed, the effect of
this reversal should be the destruction of the bond between $a$ and $b$. This means that the component
$d\bond a\bond b\bond c$ is broken into the bonds $d\bond a$ and $b\bond c$ which should
backtrack within the Petri net to capture the reversal of the transition. Nonetheless, the tokens
of $d\bond a$ must remain at place $z$. This is because a bond exists between them that has not been reversed
and was the effect of the immediately preceding transition $t_3$.
However, in the cases of $b$ and $c$ the bond can be reversed to place $y$ which is the place
where the two tokens are connected and from where they could continue to participate in any further
computation requiring their coalition. Once transition $t_2$ is subsequently reversed,
the bond between $b$ and $c$ is destroyed and thus the two tokens are able to return to their
initial places as shown in the third net in the figure. Finally, if transition $t_3$ is reversed, the bond
between $d$ and $a$ breaks and given that neither of $d$ and $a$ are connected to other elements,
the tokens can return to their initial places. From this example we observe that in
out-of-causal-order reversibility, once a transition is reversed we must reverse all the bonds that it
has created and reverse the tokens of all components created by the broken
bond as far backwards as possible.

\begin{figure}[t]
\centering
\subfigure{\includegraphics[width=.55cm]{arrow1.png}}
\subfigure{\includegraphics[width=.55cm]{arrow2.png}}
\subfigure{\includegraphics[width=.55cm]{arrow3.png}}
\subfigure{\includegraphics[width=5.4cm]{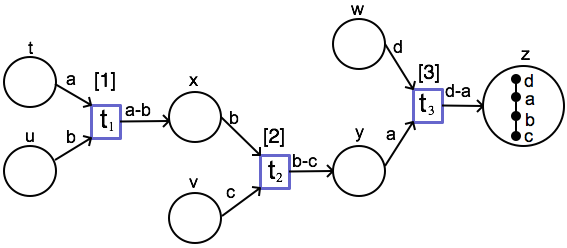}}
\subfigure{\includegraphics[width=.55cm]{arrow1r.png}}
\subfigure{\includegraphics[width=5.4cm]{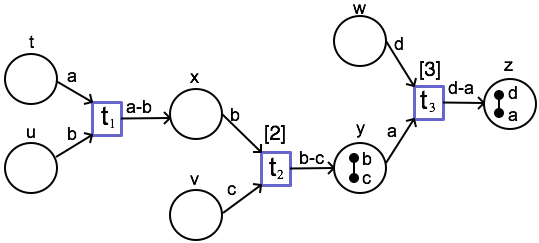}}
\hspace{.1em}
\subfigure{\includegraphics[width=.55cm]{arrow2r.png}}

\subfigure{\includegraphics[width=5.3cm]{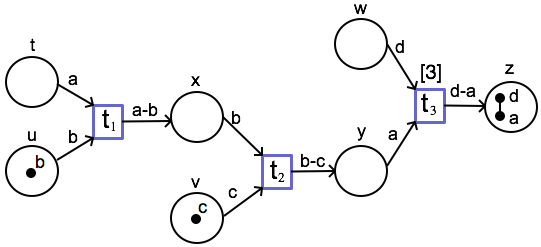}}
\subfigure{\includegraphics[width=.55cm]{arrow3r.png}}
\subfigure{\includegraphics[width=5.3cm]{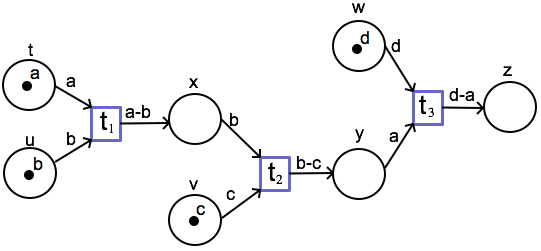}}
\setlength{\abovecaptionskip}{-1pt}
\caption{Out-of-Causal-Order example}
\end{figure}

We begin by noting that in out-of-causal order reversibility any executed transition can be reversed at any time.
\begin{definition}{\rm
Consider a \PN $(A,P,B,T,F)$, a state $\state{M}{H}$ and a transition $t\in T$. We say that $t$ is \emph{$o$-enabled} in $\state{M}{H}$, if $H(t) \in\mathbb{N}$.
}\end{definition}

The effect of reversing a transition is that all bonds created by the transition
are undone. This may result in tokens backtracking in the net. In particular,
if the reversal of a transition causes the destruction of a bond which results in a coalition of bonds 
to be broken down into
a set of subcomponents, then, each of these coalitions should
flow back, as far back as possible, to the last transition in which this sub-coalition participated. To capture the notion of ``as far backwards as possible''
we introduce the following: 

\begin{definition}\label{last}{\rm
Given a \RPN $(A,P,B,T,F)$, 
a history $H$, and 
a set of bases and bonds $C$ we write:
\[
\begin{array}{rcl}
	\first{C,H} &=& \left\{
	\begin{array}{ll}
		t , \;\;\textrm{ if }\exists t, \; \effects{t}\cap C\neq \emptyset, \;H(t) \in \mathbb{N}, \\
		 \hspace{0.2in} \not\exists t', \;\effects{t'} \cap C\neq \emptyset, \;H(t') \in \mathbb{N}, \; H(t') > H(t) \\
		\bot,  \;\textrm{ otherwise }
		\end{array}
	\right.
\end{array}
\]
}\end{definition}

Thus, $\first{C,H}$ is defined as follows: If the component $C$ has been manipulated by
some previously-executed transition,  then $\first{C,H}$ is the last executed such transition.
Otherwise, if no such transition exists (e.g. because all transitions involving $C$ have been reversed),
then $\first{C,H}$ is undefined. Transition reversal in an out-of-causal order can thus be defined as follows: 

\begin{definition}\label{oco-def}{\rm
Given a \RPN $(A, P,B,T,F)$, an initial marking $M_0$, a state $\langle M, H\rangle$ and a transition $t$ that is $o$-enabled in $\state{M}{H}$, we write $\state{M}{H}
\otrans{t} \state{M'}{H'}$
where $H'$ is defined as in Definition~\ref{br-def} and  we have:
\begin{eqnarray*}
M'(x) & = & M(x)-\effect{t}  - \;\{C_{a,x}\mid \exists a\in M(x), x\in t'\circ, t'\neq{\first{C_{a,x},  H'}}\} \\ 	 
	&& \cup \;\{C_{a,y} \mid \exists a,y,\; a\in M(y), \first{C_{a,y},H'} =t', F(t',x)\cap C_{a,y}\neq \es \} \\
 & & 			 \cup \;\{C_{a,y}\mid \exists a, y,\;a\in M(y), \first{C_{a,y},H'} =\bot,C_{a,y}\subseteq  M_0(x) \}
\end{eqnarray*}
where we use the shorthand $C_{b,z} = \connected(b,M(z)-\effect{t})$ for $b\in A$, $z\in P$.
}\end{definition}

Thus, when a transition $t$ is reversed in an out-of-order fashion all bonds that were created by
 the transition in $\effect{t}$ are undone. If the destruction of a bond divides a component 
 into smaller connected components then each of these
 components should be relocated (if needed) back to the place 
 where the complex would have existed if transition $t$ never took place, i.e., exactly after the last transition 
 that involves tokens from the sub-complex. 
 Specifically, the definition of $M'$ states that: If a token $a$ and 
 its connected components last participated in some transition with out-place $x$ other than $y$, 
 then the sub-component is removed from place $y$ and returned to place $x$, otherwise
 it is returned to the place where it occurred in the initial marking.  
 
 We may prove the following result where we write $\fotrans{}$ for $\trans{}\cup\otrans{}$. 
\begin{proposition}\label{prop5}{\rm Given a \RPN $(A, P,B,T,F)$, an initial state 
$\langle M_0, H_0\rangle$, and an execution
$\state{M_0}{H_0} \fotrans{t_1}\state{M_1}{H_1} \fotrans{t_2}\ldots \fotrans{t_n}\state{M_n}{H_n}$. The following hold:
\begin{enumerate}
\item For all $a\in A$  and $0\leq i \leq n$,   $|\{x\in P \mid a\in M_i (x)\}| = 1$, and,  if $a\in M_i(x)$ and
$t = \first{\connected(a,M_i(x)), H_i}$, if $t = \bot$ then $x$ is such that $a\in M_0(x)$, otherwise
$x\in t\circ$ and $\connected(a,M_i(x))\cap F(t,x)\neq \es$.
\item For all $\beta \in B$ and $i$, $0\leq i \leq n$, $0 \leq |\{x \in P \mid \beta\in M_i(x)\}| \leq 1$, and, 
\begin {enumerate}
\item if $t_i$ is a forward transition with $\beta\in \effect{t_i}$ then 
$\beta\in M_{i}(x)$ for some $x\in P$ and  $\beta\not\in M_{i-1}(y)$ for all $y\in P$, 
\item if $t_i$ is a reverse transition with $\beta\in \effect{t_i}$ then 
$\beta\in M_{i-1}(x)$ for some $x\in P$ and  $\beta\not\in M_{i}(y)$ for all $y\in P$, and
\item if $\beta\not\in\effect{t_i}$, $\beta\in M_{i-1}(x)$ if and only if $\beta\in M_i(x)$.
\end{enumerate}
\end{enumerate}}
\end{proposition}

 \paragraph{Proof:}
Consider a  \RPN $(A, P,B,T,F)$, an initial state 
$\langle M_0, H_0\rangle$ and an execution
$\state{M_0}{H_0} \fotrans{t_1}\state{M_1}{H_1} \fotrans{t_2}\ldots \fotrans{t_n}\state{M_n}{H_n}$. The 
proof is by induction on $n$.
\paragraph{Base Case.} For  $n=0$, by our assumption of token uniqueness, for all $a\in A$, 
$|\{x\in P \mid a\in M_0 (x)\}| = 1$, and, for all $\beta \in B$ , $0 \leq |\{x \in P \mid \beta\in M_i(x)\}| \leq 1$.
\paragraph{Induction Step.} Suppose the claim holds for all $i$, $0\leq i < n$ and consider
transition $t_n$. Two cases exist, depending on whether $t_n$ is a forward or a reverse transition:
\begin{itemize}
\item Suppose that $t_n$ is a forward transition. Then,  by Proposition~\ref{prop4} for all $a\in A$,
  $|\{x\in P \mid a\in M_n (x)\}| = 1$. Furthermore, we may show that if $a\in M_n(x)$
  either $t = \first{\connected(a,M_n(x)), H_n}$ and $x\in t\circ$, or $t= \bot$ and $x$ is such that $a\in M_0(x)$ :
  On the one hand, if $a\in\connected(b,M_{n-1}(y))$ for some $b\in F(t_n,x)$, then
  clearly $t = \first{\connected(a,M_{n-1}(y))}$ and $a\in M_n(x)$, $x\in t\circ$. On the other hand, if
  $a\not\in\connected(b,M_{n-1}(x))$ for all $b$, $y$ where $b\in F(t_n,y)$, then $a\in M_{n}(y)$ for the place $y$ such
  that  $a\in M_{n-1}(y)$
  and, by the induction hypothesis if $t=\first{\connected(a,M_n(y))}$, then clearly $t\neq t_n$ since 
  $t_n\cap \connected(a,M_{n-1}(y))= \es$, and the result follows by induction.
  Regarding clause (2), again by Proposition~\ref{prop4}, for all $\beta \in B$, $0 \leq |\{x \in P \mid \beta\in M_n(x)\}| \leq 1$,
  and if $\beta\in \effect{t_n}$ then 
$\beta\in M_{n}(x)$ for some $x\in P$ and  $\beta\not\in M_{n-1}(y)$ for all $y\in P$.

\item Suppose that $t_n$ is a reverse transition and let $a\in A$. Two cases exist:
	\begin{itemize}
	\item $a\in C_{b,x}=\connected(b,M_{n-1}(x)-\effect{t_n})$, where $\first{C_{b,x},H_n} = t$,
	and $x\not\in t\circ$ or $\first{C_{b,x},H_n} = \bot$. 
	First, let us suppose that $\first{C_{b,x},H_n} = t\neq \bot$. Note that, by the induction hypothesis, $x$ must be the unique place in
	$M_{n-1}$ with $a\in M_{n-1}(x)$. Then, by Definition~\ref{oco-def}, $a\not\in M_n(x)$ and $a\in M_n(y)$ where
	$y\in t\circ$ such that $C_{b,x}\cap F(t,y)\neq\es$. Suppose $y$ is not unique, i.e., there exists
	some $y'\neq y$ with $C_{b,x}\cap F(t,y')\neq\es$.  Further, suppose that $c\in C_{b,x}\cap F(t,y)$
	and $d\in C_{b,x}\cap F(t,y')$. By the definition of $\first{C,H}$, we may conclude that $H(t)\neq \epsilon$,
	and thus $t= t_i$ for some $i< n$. Note that during the (forward) execution of the transition, by Definition~\ref{forw}(3), 
	$c\not\in \connected(d, M_{i-1}(y))$ for all $y\in \circ t_i$. This implies that during the execution of $t_{i+1},\ldots t_n$,
	some transition(s) created a connection between $c$ and $d$, as existing in $C_{b,x}$. Thus, there must exist
	a transition $j$, $i<j\leq n$, with $(c',d')\in \effect{t}$, $c'\in \connected(c,M_{j-1}(x_{j-1}))$, $d'\in \connected(d,M_{j-1}(y_{j-1}))$,
	$x_{j-1},y_{j-1}\in \circ t_j$. However, this contradicts the choice of $t= t_i=\first{C_{b,x},H_n}$ since
	clearly $t_j$ is a transition following $t_i$ and operating on $C_{b,x}$. Consequently, we conclude that $y$ is unique.
	Thus $|\{x\in P \mid a\in M_n(x)\}| = |\{y\}= 1$ and, furthermore, $y\in t\circ$, $t=\first{C_{b,x},H_n(x)}$.
	On the other hand, if $\first{C_{b,x},H_n} = \bot$, by Definition~\ref{forw} $C\subseteq M_n(y)$, if $C\subseteq M_0(y)$.
	As before, $y$ is well defined and unique because, no transition could have created the connections of $C$
	by the definition of $\first{}$ and the assumption of token uniqueness.
	\item $a\in C_{b,x}=\connected(b,M_{n-1}(x)-\effect{t_n})$, where $\first{C_{b,x},H_n} = t$,
	and $x\in t\circ$. Note that, by the induction hypothesis, $x$ must be the unique place in
	$M_{n-1}$ with $a\in M_{n-1}(x)$. Furthermore, by Definition~\ref{oco-def}, $a\in M_n(x)$ and $a\not\in M_n(y)$
	for all $y\neq x$. 
	\end{itemize}
Now, let $\beta\in B$. If $t_n$ is a forward transition and $\beta\in\effect{t}$, the result follows from
Proposition~\ref{prop4}. If $t_n$ is a reverse transition and $\beta\in \effect{t}$, it is straightforward to see that 
$\beta\not\in M_n(x)$ for all $x$. Finally, if $\beta\not\in \effect{t}$, we may see that $\beta\in M_{n-1}(x)$ if and
only if $\beta\in M_n(x)$. This completes the proof. 
\proofend
\end{itemize}

We may now confirm that during out-of-causal-order reversing, connected components
are backtracked to the place where the components occurred as a stand-alone element
in the last state of the execution. Note that the component may have come into place various
times during the forward and backward execution of transitions and even into
various places due to possible nondeterminism in the Petri net. However, our semantics
combined with the definition of $\first{C,H,M,M_0}$ ensures that reversal of an action moves the
component to the place where it has last been used.
In what follows we write $\fotrans{}$ for $\trans{}\cup\otrans{}$. 

We begin with a useful definition.

\begin{definition}\label{c-executions}
{\rm Consider executions $\state{M_0}{H_0} \fotrans{\sigma_1} \state{M_1}{H_1}$ 
and $\state{M_0}{H_0} \fotrans{\sigma_2} \state{M_2}{H_2}$ and 
a set of bases and bonds $C=\connected(a,M_1(x)) \cap\connected(a,M_2(y))$ for some $a\in A$ and $x$, $y\in P$. We define the following:
\begin{enumerate}
\item Traces $\sigma_1$ and $\sigma_2$ are $C$-equivalent,   $\sigma_1\asymp_C \sigma_2$, if $\first{C,H_1}= \first{C,H_2}$.
\item Histories $H_1$ and $H_2$ are $C$-equivalent,  $H_1\asymp_C H_2$, if $\first{C,H_1}= \first{C,H_2}$.
\item Markings $M_1$ and $M_2$ are $C$-equivalent, $M_1\asymp_C M_2$, if $x=y$.
\item  States $\state{M_1}{H_1}$ and $\state{M_2}{H_2}$ are $C$-equivalent, $\state{M_1}{H_1}\asymp_C\state{M_2}{H_2}$, if $M_1\asymp_C M_2$ and $H_1\asymp_C H_2$.
\end{enumerate}
}\end{definition}

Thus, two traces are considered to be $C$-equivalent for a  component of tokens/bonds $C$ when they have
the same last transition manipulating $C$ (which could be undefined if none of the transitions have manipulated
$C$).  $C$-equivalence on histories is defined in the same way whereas two markings are
$C$-equivalent if they contain component $C$ in the same place. The notion is extended to states in the expected
manner.

The main result, Theorem~\ref{second} below, states  that if we have two computations beginning in the
same initial state, then they lead the complex $C$ to the same place if and only if the sequences of executed
transitions of the two computations are $C$-equivalent.

\begin{theorem}\label{second}{\rm Consider executions $\state{M_0}{H_0} \fotrans{\sigma_1} \state{M_1}{H_1}$ 
and $\state{M_0}{H_0} \fotrans{\sigma_2} \state{M_2}{H_2}$  and a complex of bonds or bases $C$. Then, 
$\sigma_1\asymp_C \sigma_2$ if and only if $\state{M_1}{H_1} \asymp_C \state{M_2}{H_2}$ 
}
\end{theorem}

\paragraph{Proof:} 
Consider sequences of transitions $\state{M_0}{H_0} \fotrans{\sigma_1} \state{M_1}{H_1}$  and $\state{M_0}{H_0} \fotrans{\sigma_2} \state{M_2}{H_2}$  and a complex of bonds and bases $C$, as 
specified by the theorem. First, let us assume that $\sigma_1\asymp_C \sigma_2$.
This implies that $\first{C,H_1} = \first{C,H_2}$ which also implies that $H_1\asymp_C H_2$. To show that $M_1\asymp_C M_2$ we
consider the following two cases:
\begin{itemize}
\item $\first{C,H_1} = \first{C,H_2}=\bot$: By Proposition~\ref{prop5},
 $C\subseteq M_1(x)$ and $C\subseteq M_2(x)$ where $C\subseteq M_0(x)$.
 This implies that $M_1\asymp_C M_2$, as required.
\item $\first{C,H_1} = \first{C,H_2} = t$: By Proposition~\ref{prop5}, for all $a\in C$, $a\in M_1(x)$, where $a\in t_1\circ$, for 
$t_1 = \first{C,H_1}$, $C\cap F(t_1,x)\neq \emptyset$ and, similarly, $a\in M_2(y)$, where $a\in t_2\circ$, for 
$t_2= \first{C,H_2}$, $C\cap F(t_2,y)\neq \emptyset$. Since $t=t_1=t_2$, $x=y$, thus $C\subseteq M_1(x)$ and $C\subseteq M_2(x)$
for some $x$, which implies that $M_1\asymp_C M_2$, as required.
\end{itemize}
For the ``only if" part of the theorem we observe that since $\state{M_1}{H_1}\asymp_C \state{M_2}{H_2}$, $H_1\asymp_C H_2$
and $\first{C,H_1} = \first{C,H_2}$. Then, by definition, $\sigma_1\asymp_C \sigma_2$, and the result follows.
\proofend

From this theorem we conclude the following corollary establishing that 
executing two causally-equivalent sequences of transitions
in the out-of-causal setting will give rise to causally equivalent states.

\begin{corollary}\label{corTheorem1}{\rm\ \  
Consider executions 
$\state{M_0}{H_0} \fotrans{\sigma_1} \state{M_1}{H_1}$  and $\state{M_0}{H_0} \fotrans{\sigma_2} \state{M_2}{H_2}$. 
If $\sigma_1\asymp\sigma_2$ then  $\state{M_1}{H_1}\asymp\state{M_2}{H_2}$.
}
\end{corollary}

\paragraph{Proof:}
Let us suppose that $\sigma_1\asymp\sigma_2$ and $\state{M_0}{H_0} \fotrans{\sigma_1} \state{M_1}{H_1}$ and
$\state{M_0}{H_0} \fotrans{\sigma_2} \state{M_2}{H_2}$. Suppose $C$ is a connected component in 
$\state{M_1}{H_1}$, i.e. $C=\connected(a,M_1(x))$ for some $x\in P$, $a\in A$. Since $\sigma_1\asymp \sigma_2$,
$H_1\asymp H_2$. Furthermore, it must be that $\sigma_1\asymp_C \sigma_2$. If not, then 
$t_1=\first{C,\sigma_1}\neq\first{C,\sigma_2}=t_2$
and $t_1$, $t_2$ concurrent actions. However, note that $t_1$ and $t_2$ are transitions both manipulating
$C$ and their effect is to forward $C$ from the inplaces to the outplaces of the transitions. 
If these transitions are concurrent, and therefore simultaneously enabled, and by the uniqueness of
tokens, it must be that they share common inplaces. Furthermore, being concurrent implies that they can be
executed in any order. For this to hold, each transition should forward $C$ back to its inplaces to allow
the other transition to fire. However, by the assumption of acyclicity of RPNs, this is not possible. Thus,
we conclude that $t_1=t_2$ which implies that $\sigma_1\asymp_C\sigma_2$ and, by Theorem~\ref{second},
$C\subseteq M_2(x)$. Since this argument holds for any component $C$, we deduce that $M_1=M_2$ and
the result follows

In addition, the following corollary establishes that out-of-causal-order reversibility is consistent with
standard forward execution in the sense that out-of-causal reversibility will never return tokens to places that are 
unreachable during forward execution. 

\begin{corollary} \label{corForward}{\rm\ \
Consider executions $\state{M_0}{H_0} \fotrans{\sigma_1} \state{M_1}{H_1}$ and
$\state{M_0}{H_0} \fotrans{\sigma_2} \state{M_2}{H_2}$  where
$\sigma_1$ is a trace with both forward and backward transitions and $\sigma_2$ is a trace with only forward
transitions and $\sigma_1 \asymp_C \sigma_2$. 
Then, for $x\in P$, $C\subseteq M_1(x)$ if and only if $C\subseteq M_2(x)$.
}
\end{corollary}

\paragraph{Proof:}
From Theorem~\ref{second} we know that since $\sigma_1\asymp_C \sigma_2$, $M_1\asymp_C M_2$. Thus,
$C\subseteq M_1(x)$, implies that $C\subseteq M_2(x)$ and the result follows.
 \proofend

Finally, we state the following result that demonstrates the relation between the three forms of
reversibility, as proposed for RPN's.
\begin{proposition}\label{connect}{\rm\ \
$\btrans{} \subset\ctrans{}\subset \otrans{}$. 
}
\end{proposition}

\paragraph{Proof:}
To prove the proposition consider a \RPN $(A, P,B,T,F)$, a state $\langle M, H\rangle$ and suppose
that transition $t$ is $bt$-enabled and $\langle M, H\rangle\btrans{t} \langle M', H'\rangle$. Then,
by definition of $bt$-enabledness, $H(t)>H(t')$ for all $t'\neq t$, $H(t)\in \mathbb{N}$. This implies
that $t$ is also $co$-enabled, and by the definition of $\ctrans{}$, we conclude that 
 $\langle M, H\rangle\btrans{t} \langle M', H'\rangle$. It is easy to see that the inclusion is strict,
 as for example illustrated in Figure~4.

For the second inclusion, let us suppose that transition $t$ is $co$-enabled and $\langle M, H\rangle\ctrans{t}
\langle M_1, H_1\rangle$. Then,
by the definition of $o$-enabledness, $t$ is $o$-enabled. Suppose $\langle M, H\rangle\otrans{t}\langle M_2, H_2\rangle$.
It is easy to see that in fact $H_1 = H_2$ (the two histories are as $H$ with the exception that $H_1(t) = H_2(t) = \epsilon$).
It remains to show that $M_1 = M_2$.  

Let $a\in A$. We must show that
$a\in M_1(x)$ if and only if $a\in M_2(x)$. Two cases exist:
\begin{itemize}
\item If $a\not\in \connected(b,M(y))$ for all $b\in A$, $y\in t\circ$, then $a\in M_1(x)$ and $a\in M_2(x)$ such that $a\in M(x)$.
\item If $a\in \connected(b,M(y))$ for some $b\in A$, $y\in t\circ$, then
consider $x\in \circ t$ with $\connected(a,M(y))\cap F(x,t)\neq \es$. Such a place exists, by Definition~\ref{well-formed}(3),
and it is unique, by the assumption of uniqueness of tokens. Note that by Definition~\ref{br-def}, $a\in M_1(x)$. 
Consider $t'$ such that $x\in t'\circ$ and $F(t',x)\cap \connected(a,M(y))\neq \es$. If such a transition exists
 by observing that $t'$ would not have been $co$-enabled
since $H(t)\neq \epsilon$ and $t'$ and $t$ manipulate common tokens in $\connected(a, M(y))$,
we conclude that $\first{\connected(a,M(y)),H}=t'$.
As a result  $a\in M_2(x)$, as required. 
On the other hand, if such a $t'$ does not exist, then it must be that $\first{\connected(a,M(y)),H}=\bot$  and,
by the definition of $\otrans{}$, $a\in M_2(x)$ where 
$a\in M(x)$ as required.
\end{itemize}
Now let $\beta\in B$. We must show that
$\beta\in M_1(x)$ if and only if $\beta\in M_2(x)$. Two cases exist:
\begin{itemize}
\item If $\beta\not\in \connected(b,M(y))$ for all $b\in A$, $y\in t\circ$, then $\beta\in M_1(x)$ if and only if  
$\beta\in M(x)$ if and only if $\beta\in M_2(x)$.
\item If $\beta\in \connected(b,M(y))$ for some $b\in A$, $y\in t\circ$, then, if also $\beta\in\effect{t}$, then
$\beta\not\in M_1(x)$ for all $x\in P$ and, similarly, for $M_2$. If, however, $\beta\not\in\effect{t}$, then
consider $x\in \circ t$ with $\connected(a,M(y)-\effect{t})\cap F(x,t)\neq \es$. Such a place exists, by Definition~\ref{well-formed}(3),
and it is unique, by the assumption of uniqueness of tokens. Note that by Definition~\ref{br-def}, $\beta\in M_1(x)$. 
Consider $t'$ such that $x\in t'\circ$ and $F(t',x)\cap \connected(a,M(y)-\effect{t})\neq \es$. If such a transition exists
 by observing that $t'$ would not have been $co$-enabled
since $H(t)\neq \epsilon$ and $t'$ and $t$ manipulate common tokens in $\connected(a, M(y)-\effect{t})$,
we conclude that $\first{\connected(a,M(y)-\effect{t}),H}=t'$.
As a result  $a\in M_2(x)$, as required. 
On the other hand, if such a $t'$ does not exist, then it must be that $\first{\connected(a,M(y)-\effect{t}),H}=\bot$  and,
by the definition of $\otrans{}$, $a\in M_2(x)$ where 
$a\in M(x)$ as required.
\end{itemize}
 Again, we may observe that the inclusion is strict given that out-of-causal reversal allows to reverse all executed transitions
 and not only those whose effects have been undone.
This completes the proof.
\proofend

\begin{figure}[t]
\centering
\subfigure{\includegraphics[width=.55cm]{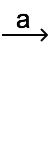}}
\subfigure{\includegraphics[width=.55cm]{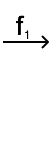}}
\subfigure{\includegraphics[width=.55cm]{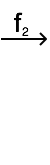}}
\subfigure{\includegraphics[width=7cm]{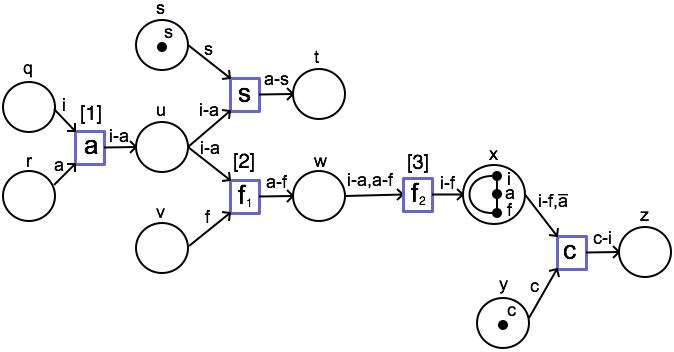}}
\subfigure{\includegraphics[width=.55cm]{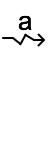}}
\subfigure{\includegraphics[width=7cm]{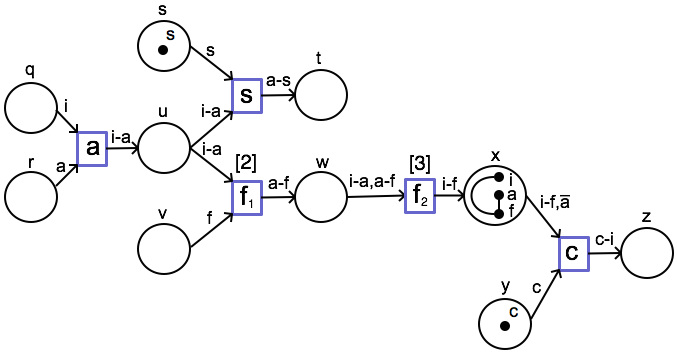}}
\hspace{.1em}
\subfigure{\includegraphics[width=.55cm]{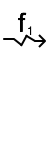}}
\subfigure{\includegraphics[width=7cm]{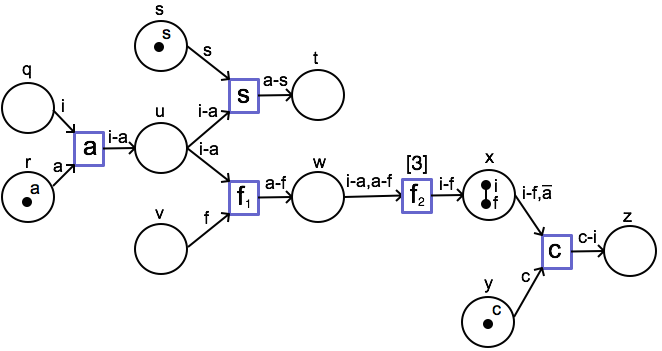}}
\subfigure{\includegraphics[width=.55cm]{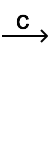}}
\subfigure{\includegraphics[width=7cm]{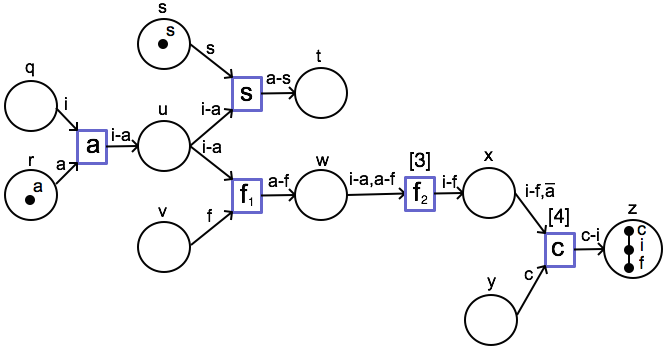}}
\subfigure{\includegraphics[width=.55cm]{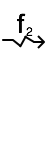}}
\subfigure{\includegraphics[width=7cm]{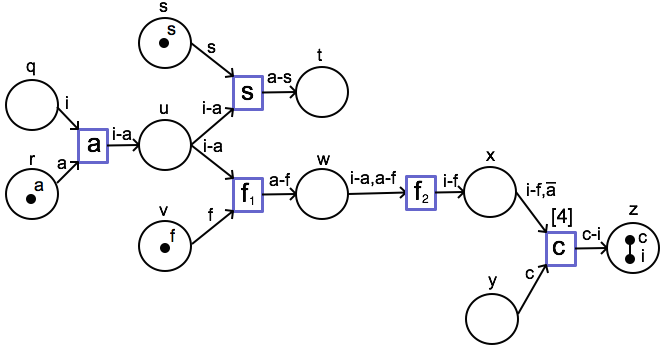}}
\setlength{\abovecaptionskip}{-1pt}
\caption{Transaction processing}\label{o-example}
\label{trans}
\end{figure}

In the following
example we illustrate how reversing Petri nets featuring out-of-causal-order reversibility can be used for modelling a simplified
transactional system.
\begin{example}{\rm\ \
Transaction processing manages sequences of operations, also called transactions, that can either succeed 
or fail as a complete unit. Specifically, a long-running transaction consists of a sequence of steps.
Each of these steps may either succeed, in which case the flow of control moves on to the next atomic 
step in the sequence, or it may
fail, in which case a compensating transaction is often used to undo failed transactions and restore the 
system to a previous state.   
If all steps of the transaction execute successfully then the transaction is considered as successful and
it is committed.

In Figure~\ref{trans} we consider a model of such a transaction. Due to the size of the 
net we restrict our attention to a transaction
with only one step which nonetheless illustrates sufficiently the mechanisms needed to model
the system. The intuition
is as follows: for the execution of the transaction to commence it is necessary for token $i$ to be available. This token
is bonded with token $a$ in which case transition $a$ can be executed with the effect of creating the bond $i\bond a$ in place $u$.
At this stage there are two possible continuations. The first possibility is that the bond $i\bond a$ will participate in transition $s$ which 
models the successful completion of  step $a$ as well as the transaction, yielding the bond $i\bond a\bond s$. The second possibility is
that step $a$ fails. In this case, token $f$ comes in place and the failure is implemented via transitions $f_1$ and $f_2$ as follows: 
To begin with in action $f_1$, token $f$ is bonded with token $a$, whereas in action $f_2$ token $i$ is bonded
with token $f$. At this stage the compensation comes in place (token $c$) where the intention is that step $a$ should be undone.
In our model, this involves undoing transition $a$. Note that this will have to be done according to  our out-of-causal-order
definition since transition $a$ was followed by $f_1$ and $f_2$ which have not been undone. Only once this is accomplished,
will the precondition of transition $c$, namely $\overline{a}$, be enabled. In this case, transition $c$ can be executed leading
to the creation of bond $i\bond c$ in place $z$.

Figure~\ref{trans} shows a possible execution of the system beginning at the state where the sequence of transitions $a;f_1;f_2$ has
already taken place. Next, computation can not go forward as already explained, but, as illustrated in the first transition in the figure,
transition $a$ may be reversed, causing the destruction of  bond $i\bond a$. Next, transition $f_1$ is
reversed, causing the destruction of bond $a\bond f$, in which case token $a$ can return to its initial place, capturing that
the step has been reversed. At this stage transition $c$ can be executed in the forward direction leading to the execution of
the compensation. In the final step, $f_2$ is reversed causing the reversal of the bond $i\bond f$ and recovering the failure token
back to its initial place.

}\end{example}


\section{Conclusions}\label{sec:Conclusions}

This paper proposes a reversible approach to Petri nets that allows the modelling
of reversibility as realised by backtracking, causal reversing and out-of-causal-order reversing. 
To the best of our knowledge, this is the first such proposal, since the only related
previous work of~\cite{PetriNets,BoundedPNs}, having a different aim, implemented a very liberal
way of reversing computation in Petri nets by introducing additional reversed transitions.
Indeed, our proposal allows systems to reverse at any time 
leading to previously visited states or even to new ones without the need of additional forward actions. 
Moreover, this  interpretation of Petri nets  has the capability for reversing without the need of an
extensive memory.

Indeed, good models that can be easily understood and simulated, even by scientists with expertise outside
Computer Science,
can prove very useful to understand complex systems. In particular, the expressive power and visual nature
offered by Petri nets coupled with reversible computation has the potential of providing an attractive setting for
studying, analysing, or even imagining alternatives in a wide range of systems.

\remove{In our current research we are working on relaxing the restrictions we imposed in the Petri net model of the
present work. Immediate extensions include allowing multiple tokens of the same base/type to occur in
a model, as well as the addition of cycles. Furthermore, we are currently working towards defining causal forms
of reversibility in the model 
of~\cite{GlabbeekGS11} where additional memory is needed to store the sequence of transition execution.  
Our research will continue by considering out-of-causal-order reversibility in the general model where we 
expect that original mechanisms will be required for allowing to capture the reversal of transitions in an 
out-of-causal order.

As future work, we are planning to extend our formalism by considering approaches for controlling reversibility,
as for instance
in~\cite{TransactionsRCCS,ERK,LaneseMSS11}. In fact, we are currently exploring this direction with the use of
probabilities that can capture the likelihood of a transition executing in the forward or backward 
irection~\cite{Statistical}. In addition, it would be interesting to provide models where transitions
are associated with a ``reversibility type'' that specifies whether a transition can be reversed in a causal or
out-of-causal order. Furthermore, we would like
to further apply our framework in the fields of biochemistry and long-running transactions. Finally, we intend to
investigate the expressiveness relationship between Reversible Petri Nets and standard Petri nets and explore
how reversible computation affects the expressive power of various subclasses of Petri nets.
}

In our current research we are investigating the expressiveness relationship
between RPNs and coloured PNs where we expect that the global control implemented
by histories can be encoded in coloured PNs using additional transitions and places. We aim
to provide and prove the correctness of such a translation and analyse the
associated trade-off in terms of Petri net size.
Furthermore, we are working on relaxing the restrictions we imposed in the RPN model of the
present work such as 
allowing multiple tokens of the same base/type to occur in
a model, as well as the addition of cycles. Note that
the addition of cycles can in fact be achieved within our model by adopting
histories in the form of a stack for each transition that recalls all previous occurrences of the transition.

 As future work, we are planning to extend our formalism by considering approaches for controlling reversibility,
as for instance
in~\cite{TransactionsRCCS,ERK,LaneseMSS11}. We plan to explore this direction with the use of
probabilities that can capture the likelihood of a transition executing in the forward or backward 
direction~\cite{Statistical}. 
Finally, we would like
to further apply our framework in the fields of biochemistry and long-running transactions.

\noindent{\bf{Acknowledgents:}} This research was partially supported by the EU COST Action IC1405.
We are grateful to K. Barylska,  A. Gogolinska, L. Mikulski, and M. Piatkowski  for interesting discussions on previous drafts of this work.

\bibliographystyle{abbrv}
\bibliography{References}
\newpage

\end{document}